%% file: paper.tex
\documentclass[12pt,a4paper]{article}
\usepackage{a4wide}
\usepackage{graphicx}
\usepackage[centertags]{amsmath}
\usepackage{amssymb}
\usepackage{cite}
\usepackage{epsfig}
\usepackage{longtable}
\usepackage{enumerate}
\usepackage{scrtime}
\usepackage{color}
\usepackage[sans]{dsfont} 
\usepackage{textcomp}
\usepackage{pifont}
\usepackage{verbatim}
\usepackage{ulem}
\usepackage{colortbl}
\usepackage{epic}
\usepackage{subfigure}
\usepackage{paralist}

\usepackage{pstricks}
\usepackage{pst-node}

\psset{linearc=0.15}

\newpsstyle{boxstyle}{framearc=0.2,fillstyle=solid,fillcolor=lightgray,framesep=4mm}
\newsavebox\PBox
\def\myBox#1#2#3{%
 \sbox\PBox{\psframebox[style=boxstyle]{\parbox{#1}{#3}}}%
 \begin{pspicture}(0,-\ht\PBox)(\wd\PBox,1.2\ht\PBox)
   \rput[l](0,0){\usebox\PBox}
   \rput[l](5\fboxsep,\ht\PBox){\colorbox{yellow}{#2\hspace{\fboxsep}}}
 \end{pspicture}%
}

\newcommand{\bs}[1]{\boldsymbol{#1}}
\newcommand{\bsb}[1]{\boldsymbol{\overline{#1}}}
\newcommand{\ra}{\rightarrow}

\newcommand{\Hc}{\bar{H}}

\newcommand{\uc}{\bar{u}}
\newcommand{\dc}{\bar{d}}
\newcommand{\ec}{\bar{e}}
\newcommand{\nc}{\bar{\nu}}
\newcommand{\m}{\text{-}}

\newcommand{\vectornorm}[1]{\left|\left|#1\right|\right|}

\graphicspath{{./}{./graphs/}}

\begin{document}

\thispagestyle{empty}

\begin{flushright}
OHSTPY-HEP-T-07-002\\
\end{flushright}
\vskip 2cm

\begin{center}
{\huge Can String Theory\\[1ex]
Predict the Weinberg Angle?}
\vspace*{5mm} \vspace*{1cm}
\end{center}
\vspace*{5mm} \noindent
\vskip 0.5cm
\centerline{\bf Stuart Raby, Ak\i{}n Wingerter}
\vskip 1cm
\centerline{
\em Department of Physics, The Ohio State University,}
\centerline{\em 191 W.~Woodruff Ave, Columbus, OH 43210, USA}
\vskip2cm

\centerline{\bf Abstract}
\vskip .3cm

We investigate whether the hypercharge assignments in the Standard
Model can be interpreted as a hint at Grand Unification in the
context of heterotic string theory. To this end, we introduce a
general method to calculate $\text{U}(1)_Y$ for any heterotic
orbifold and compare our findings to the cases where hypercharge
arises from a \textsc{gut}. Surprisingly, in the overwhelming
majority of 3-2 Standard Models, a non-anomalous hypercharge
direction can be defined, for which the spectrum is vector-like. For
these models, we calculate $\sin^2 \theta_w$ to see how well it
agrees with the standard \textsc{gut} value.  We find that 12\%
have $\sin^2\theta_w = 3/8$, while all others have values which are less.
Finally, 89\% of the models with $\sin^2\theta_w = 3/8$ have
$\text{U}(1)_Y \subset \text{SU}(5)$.

\vskip .3cm

\newpage


\section{Introduction}

\begin{figure}[h!]
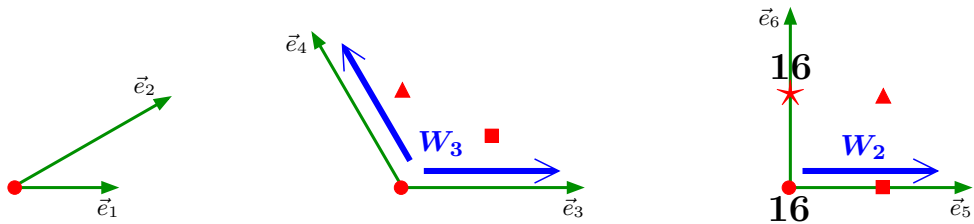

\centering
\begin{center}\input ./graphs/firsttwisted.pstex_t \end{center}
\caption{The first twisted sector of the
$\mathbb{Z}_6$-II orbifold. The shift vectors of the mini-landscape search preserve a ``local" $\text{SO}(10)$ (or
$\text{E}_6$) \textsc{gut} at the fixed points $(\color{red}{\bullet},\color{red}{\bullet},\color{red}{\bullet})$ and $(\color{red}{\bullet},\color{red}{\bullet},\color{red}{\star})$ with two $\bs{16}$ (or $\bs{27}$)
dimensional massless multiplets.}
\label{fig:orbifold}
\end{figure}

In a recent paper, Lebedev et al. \cite{Lebedev:2006kn} performed a
``mini-landscape" search in the heterotic string looking for
\textsc{mssm}-like models. The search focused on the $\mathbb{Z}_6$-II orbifold and
quite dramatically it was found that about 1 in 300 theories were
\textsc{mssm}-like.  Why should there be such a ``fertile patch" in the
string landscape can be understood in terms of the property of
orbifold \textsc{gut}s and ``local \textsc{gut}s" as emphasized in the following Refs.
\cite{Kobayashi:2004ud,Forste:2004ie,Kobayashi:2004ya,Buchmuller:2006ik,Buchmuller:2004hv,Buchmuller:2005jr,Buchmuller:2005sh}.
In the $\mathbb{Z}_6$-II orbifold of the heterotic string,  four particular
shift vectors were identified that break the visible $\text{E}_8$ gauge
symmetry to $\text{E}_6$ or $\text{SO}(10)$.   Then in the first twisted sector
(see Fig. \ref{fig:orbifold}) it was shown that massless chiral
multiplets in the {\bf 27} of $\text{E}_6$ or {\bf 16} of $\text{SO}(10)$ exist,
residing at ``local" \textsc{gut} twisted sector fixed points. In fact, with
one Wilson line lying in the $\text{SO}(4)$ torus, there are exactly two
possible families of quarks and leptons.   With two Wilson lines, it
was possible to break the visible $\text{E}_8$ to the Standard Model
($\text{SU}(3) \times \text{SU}(2) \times \text{U}(1)_Y \times $ a hidden sector gauge
symmetry). Two families were localized at the ``local" \textsc{gut} fixed
points and the third family and Higgs doublets were found in a
combination of the untwisted and twisted sectors.   In the
``mini-landscape" search \cite{Lebedev:2006kn}, the SM (including
hypercharge) was constrained to lie within $\text{SU}(5) \subset \text{SO}(10) \;
{\rm or} \; \text{E}_6$.   This short review describes the ``fertile patch"
in the landscape and hopefully makes it clear why it was so fertile.

In the present paper we also focus on the $\mathbb{Z}_6$-II orbifold with the
same shifts.   We inherit the same $\text{SO}(10)$ symmetry at the two
$\text{SO}(4)$ fixed points.   Our search differs from Ref.
\cite{Lebedev:2006kn} in that {\it we do not demand} that the SM sit
within $\text{SU}(5) \subset \text{SO}(10)$.   Our search is more general, so that
the fundamental question we address is - to what extent do \textsc{mssm}-like
models require an intermediate \textsc{gut}?    In the following Section
\ref{sec:hypercharge_from_guts} we refine this question.  In Section
\ref{sec:hypercharge_general} we outline our more general search for
\textsc{mssm}-like models.  Then in Section \ref{sec:results} we discuss our
results, including a comparison to the mini-landscape analysis.


\section{Hypercharge from Grand Unification}
\label{sec:hypercharge_from_guts}

In string compactifications aiming at the Standard Model in four dimensions, one is typically
left with a large number of abelian factors\footnote{In (heterotic) orbifold constructions,
$\text{U}(1)$ factors may be broken by the mechanism of {\it continuous Wilson lines}, as
explored in Refs.~\cite{Forste:2005rs, Forste:2005gc}.},
\begin{equation}
\text{SU}(3) \times \text{SU}(2) \times \text{U}(1)^n \times \text{non-abelian factors}.
\end{equation}
A priori, hypercharge may be {\it any} linear combination of the
$\text{U}(1)$'s that gives the correct values on the elementary
particles, cf.~Tab.~\ref{tab:spectrum_SM}. What complicates matters
even more is the fact that generically, the spectrum contains a
large number of representations that may play the roles of quarks,
leptons, and Higgses in the low-energy theory. In order not to be in
gross contradiction to experimental observation, the spectrum is
required to be {\it vector-like} so that the exotic
particles\footnote{All the additional particles that are not in the
\textsc{mssm} are termed {\it exotics}. If the exotics are such that
they can be combined to pairs of representations that are
conjugates of each other, the spectrum is called {\it vector-like}.}
may decouple by acquiring a large Majorana-type mass. In this
context, it is not clear which representations are to be identified
with families or exotics, respectively. As a result, we are
left with a large number of choices, which makes finding a $\text{U}(1)$ which may
play the role of hypercharge a hard problem to solve.

\begin{table}[h!]
\centering
\normalsize
\renewcommand{\arraystretch}{1.4}
\begin{tabular}{|c|l||c|l||c|l|}
\hline
$Q$ & $(\bs{3}, \bs{2})_{1/3}$       &   $L$ & $(\bs{1}, \bs{2})_{\m 1}$   &  $H$ & $(\bs{1}, \bs{2})_{1}$     \\
$\uc$ & $(\bsb{3}, \bs{1})_{\m 4/3}$ &   $\ec$ & $(\bs{1}, \bs{1})_{2}$    &  $\Hc$ & $(\bs{1}, \bs{2})_{\m 1}$\\
\cline{5-6}
$\dc$ & $(\bsb{3}, \bs{1})_{2/3}$    &   $\nc$ & $(\bs{1}, \bs{1})_{0}$    &  \multicolumn{2}{|c|}{}            \\
\hline
\end{tabular}
\caption{Matter content of the Standard Model. In our conventions, ${\displaystyle Q = T_{3L} + \dfrac{Y}{2}}$.}
\label{tab:spectrum_SM}
\end{table}

\begin{figure}[h!]
\centering
\begin{tabular}{ccccc}
{
\unitlength=.8mm
\begin{picture}(47,21)
\thicklines
\put(2, 5){\circle{4}} \put(17,5){\circle{4}} \put(32,5){\circle{4}} \put(32,20){\circle{4}} \put(47,5){\circle{4}}
\put(4,5){\line(1,0){11}} \put(19,5){\line(1,0){11}} \put(32,7){\line(0,1){11}} \put(34,5){\line(1,0){11}}
\put(0,-1){$\alpha_1$} \put(15,-1){$\alpha_2$} \put(30,-1){$\alpha_3$} \put(35,19){$\alpha_5$} \put(45,-1){$\alpha_4$}
\end{picture}
} &  \raisebox{4ex}[-3ex]{ $\bs{\longrightarrow}$ } & {
\unitlength=.8mm
\begin{picture}(47,21)
\thicklines
\put(2, 5){\circle{4}} \put(17,5){\circle{4}} \put(32,5){\circle{4}} \put(32,20){\circle{4}} \put(47,5){\circle{4}}
\put(4,5){\line(1,0){11}} \put(19,5){\line(1,0){11}} \put(32,7){\line(0,1){11}} \put(34,5){\line(1,0){11}}
\put(0,-1){$\alpha_1$} \put(15,-1){$\alpha_2$} \put(30,-1){$\alpha_3$} \put(35,19){$\alpha_5$} \put(45,-1){$\alpha_4$}
\put(29,17){\line(1,1){6}} \put(29,23){\line(1,-1){6}}
\end{picture}
} &  \raisebox{4ex}[-3ex]{ $\bs{\longrightarrow}$  \hspace{-4ex}} & {
\unitlength=.8mm
\begin{picture}(47,21)
\thicklines
\put(2, 5){\circle{4}} \put(17,5){\circle{4}} \put(32,5){\circle{4}} \put(32,20){\circle{4}} \put(47,5){\circle{4}}
\put(4,5){\line(1,0){11}} \put(19,5){\line(1,0){11}} \put(32,7){\line(0,1){11}} \put(34,5){\line(1,0){11}}
\put(0,-1){$\alpha_1$} \put(15,-1){$\alpha_2$} \put(30,-1){$\alpha_3$} \put(35,19){$\alpha_5$} \put(45,-1){$\alpha_4$}
\put(29,2){\line(1,1){6}} \put(29,8){\line(1,-1){6}}
\put(29,17){\line(1,1){6}} \put(29,23){\line(1,-1){6}}
\end{picture}
}\\[3ex]
$\text{SO}(10)$ &  & $\text{SU}(5) \times \text{U}(1)_X$  &  & $\text{SU}(3)_c \times \text{SU}(2)_L \times
\text{U}(1)_X \times \text{U}(1)_Y$
\end{tabular}
\caption{From $\text{SO}(10)$ to (un)flipped $\text{SU}(5)$ to the Standard Model.}
\label{fig:SO10_to_SU5_to_SM}
\end{figure}
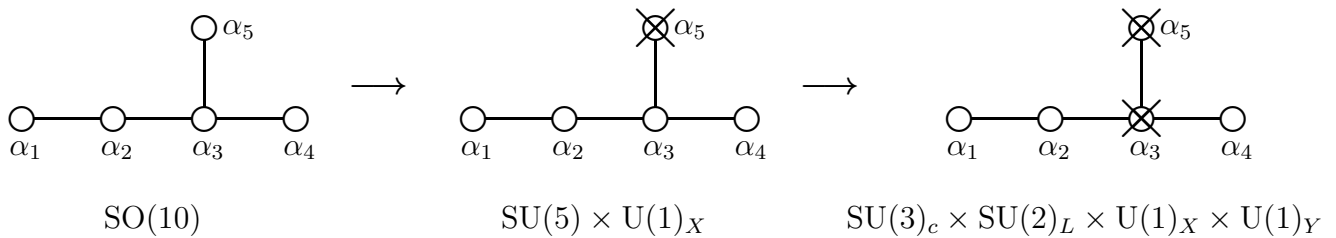

The problem is not so difficult to solve, if we assume some additional structure. Grand Unification
\cite{Georgi:1974sy, Pati:1974yy, Fritzsch:1974nn} offers a default choice for hypercharge. In
Georgi-Glashow $\text{SU}(5)$ \cite{Georgi:1974sy}, $\text{U}(1)_Y$ is obtained as the abelian
factor that is left over after breaking $\text{SU}(5)$ to the Standard Model gauge group, see
Fig.~\ref{fig:SO10_to_SU5_to_SM}. More specifically, the hypercharge direction \cite{Slansky:1981yr}
is given by the dual of the root that is projected out:
\begin{equation}
Y_{\text{GG}} = \frac{5}{3} \alpha_3^* = \sum_{j} \frac{5}{3} \, (A_{\text{SU}(5)}^{-1})_{3j} \,
\alpha_j = \frac{1}{3} \left( 2\alpha_1 + 4\alpha_2 + 6\alpha_3 + 3\alpha_4 \right)
\end{equation}
This works not only in the case where we start with a grand unified gauge group in four dimensions,
but also in any other construction where the Standard Model is realized as a subset of a grand
unified group. A particular example of this type is a field theoretic or stringy orbifold where
Grand Unification is realized in higher dimensions \cite{Asaka:2001eh, Forste:2004ie}.

In a recent publication \cite{Lebedev:2006kn}, we exploited this very fact\footnote{The $\text{SO}(10)$
structure was required not only for hypercharge, but also for a couple of other reasons, such as gauge
and Yukawa coupling unification, the presence of a $B-L$ symmetry, a discrete subgroup of which may be
a candidate for $R$-parity, etc. A nice discussion of the role of $\text{SO}(10)$ in the context of
theories beyond the Standard Model can be found in Refs.~\cite{Dimopoulos:1991au, Nilles:2004ej}. }
to derive 223 heterotic $\mathbb{Z}_6$-II orbifold models with three generations of quarks and leptons,
one pair of Higgses, and a vector-like spectrum. The first step in the analysis was a judicious choice
of a shift vector leading to either $\text{SO}(10)$ or $\text{E}_6$, with the additional requirement
that at least one family representation, $\bs{16}$ or $\bs{27}$, respectively, be localized in the
first twisted sector. Wilson lines subsequently broke the gauge symmetry to that of the Standard Model
 and hypercharge was realized as the $\text{U}(1)$ which lies in $\text{SU}(5)$, cf.~again
Fig.~\ref{fig:SO10_to_SU5_to_SM}.

\begin{figure}[h]
\centering \epsfig{figure=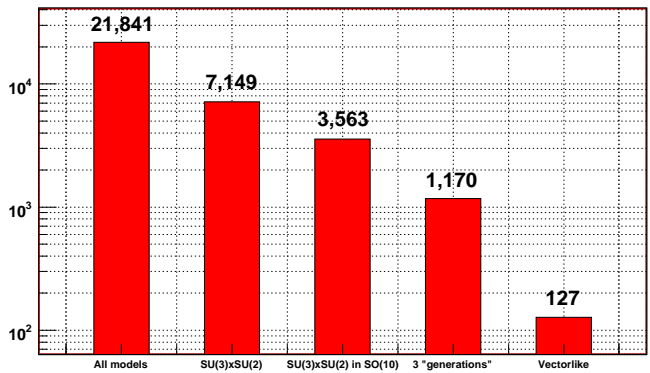, angle=0,
scale=0.5} \caption{Consider shift $V^{\text{SO}(10),1}$ of
Ref.~\cite{Lebedev:2006kn}. Of the (i) 21,841 models, (ii) 33\%
contain the SM gauge group in at least one $\text{E}_8$, (iii) 16\%
contain the SM gauge group as a subset of $\text{SO}(10)$, (iv) 5\%
are three-generation Standard Models (w.r.t.~color and weak isospin)
lying in $\text{SO}(10)$, and (v) 0.6\% are three-generation
Standard Models (now including hypercharge) with a vector-like
spectrum where hypercharge is that of Georgi-Glashow $\text{SU}(5)
\subset \text{SO}(10)$.} \label{fig:models_SO10_hypercharge}
\end{figure}

In light of the success of this approach, one question immediately
arises: How crucial was the assumption that the Standard Model gauge
group be realized as a subgroup of $\text{SO}(10)$? If we drop this
assumption, can we still find a suitable $\text{U}(1)$ direction
that plays the role of hypercharge such that all family and Higgs
representations have the standard assignments and the remaining
particles can be combined to vector-like pairs that decouple from
the low-energy theory?

\medskip

In Section \ref{sec:hypercharge_general}, we present a
general method for constructing a suitable hypercharge out of any
number of $\text{U}(1)$ factors. Although we specialize to a heterotic
orbifold model, it should be clear that the outlined methods are applicable
in a much broader context.


\section{Searching for Hypercharge in the General Case}
\label{sec:hypercharge_general}

For the sake of clarity, we will consider a specific example along with the general construction.
Tab.~\ref{tab:spectrum_model_yes_1} shows the spectrum of a typical $\mathbb{Z}_6$-II orbifold
model (More details are given in Appendix \ref{app:details_model_yes_1}).

\begin{table}[h!]
\centering
\normalsize
\renewcommand{\arraystretch}{1.4}
\begin{tabular}{|r|r|r|r|}
\hline
$  3  \times (  \bs{3},  \bs{2},  \bs{1})$ &   $ 12  \times ( \bsb{3},  \bs{1},  \bs{1})$ &   $ 29  \times (  \bs{1},  \bs{2},  \bs{1})$ &   $ 8  \times (  \bs{1},  \bs{1}, \bs{5})$ \\
\hline
                                           &   $ 6  \times (  \bs{3},  \bs{1},  \bs{1})$ &    $136  \times (  \bs{1},  \bs{1},  \bs{1})$ &   $ 8  \times (  \bs{1},  \bs{1},  \bsb{5})$ \\
\hline
\end{tabular}
\caption{Three-generation model with gauge group $\text{SU}(3)\times\text{SU}(2)\times \text{U}(1)^5 \times
\text{SU}(5)' \times {\text{U}(1)'}^4$.}
\label{tab:spectrum_model_yes_1}
\end{table}

In the present case, hypercharge may be any linear combination of 9 abelian factors,
\begin{equation}
\text{U}(1)_Y = x_1 \text{U}(1)_1 + x_2 \text{U}(1)_2 + \ldots +  x_9 \text{U}(1)_9.
\label{eq:ansatz_for_hypercharge}
\end{equation}

The naive ansatz of choosing a subset of the representations in Tab.~\ref{tab:spectrum_model_yes_1}
to be the observed elementary particles and then solving the linear equations for their having the
correct hypercharges is not feasible. The number of such choices\footnote{The model under consideration
is relatively well-behaved with regard to the multiplicities of particles. There exist cases in which
the net number of generations is realized by differences of larger multiplicities, thus considerably
increasing the complexity of the problem.} is at least
\begin{equation}
{3 \choose 3} \times {12 \choose 6} \times {6 \choose 3} \times {29 \choose 5} \times {5 \choose 4}
\times {216 \choose 3} \times {216 \choose 3} \quad > \quad 3 \times 10^{22}
\end{equation}
corresponding to choosing (i) left-handed quark doublets, (ii) right-handed quarks, (iii) distinguishing
up- and down-type quarks, (iv) 3 lepton doublets and 2 Higgses, (v) distinguishing leptons and up-type
Higgs, (vi) right-handed electrons, (vii) right-handed neutrinos. (Note that in the above counting, the charges under
the hidden sector gauge group are multiplicities for the Standard Model particles.)

\medskip

Even if we were able to cope with such a large number of cases, the solutions to the linear equations
will in general not give a vector-like spectrum. From these considerations it should be clear that
another line of thought has to be pursued.

\medskip

In the remainder of this section, we describe our strategy for finding a suitable hypercharge even under
the most unfavorable circumstances. For facility of inspection, in the following we will enumerate the
key ideas entering our calculations.

\subsection{Exploiting the Linear Constraints}
\label{sec:linear-constraints}

\begin{inparaenum}[(i)]
\item \label{it:i}A judicious choice for the basis of $\text{U}(1)$ directions greatly simplifies the calculations.
In principle, one may choose any $n$ linearly independent directions that are orthogonal
to the simple roots of the unbroken gauge group. However, from experience with Grand Unified Theories {\it we expect}
the $\text{U}(1)$ directions (up to some multiplicative factor) to correspond to the duals of the roots
that are projected out, so we can take them to lie in the root lattice\footnote{Observe that the root
lattice of $\text{E}_8\times \text{E}_8'$ is self-dual.} of $\text{E}_8 \times \text{E}_8'$. This is not
only, in some sense, the default choice, but it also then guarantees that $N$-times the $\text{U}(1)$ directions
will have integer values\footnote{The highest weight of a representation is of the general form $p+k V$,
where $p$ is in the $\text{E}_8\times \text{E}_8'$ root lattice, $k$ the number of the twisted sector, and
$V$ the shift vector. Our assertion follows from the fact that $N V$ is in the root lattice.\label{footnote:multiply_by_N}}
on the representations, where we denote by $N$ the order of the orbifold.

\medskip

\item Unfortunately, this assumption is not fully justified. Although it is generally true that we may
always find $n$ linearly independent basis vectors that span the vector space orthogonal to the simple
roots and at the same time belong to the $\text{E}_8 \times \text{E}_8'$ lattice, it is a priori not clear
why hypercharge should lie in the lattice, or in other words, why the coefficients in
Eq.~(\ref{eq:hypercharge_direction}) below should be integers. (The deeper reason behind this is the
existence of chiral exotics which may have hypercharge assignments that are not multiples of 1/3.)
Therefore, we will drop the assumption on the coefficients when we consider the general case in
Section \ref{sec:non-linear-constraints}.

\medskip

\item Since assigning the correct hypercharges to all family representations has proved to be
impractical, we make a compromise and demand only that the left-handed quark doublets and the
right-handed quark singlets have the correct values. In this case, there is still a choice to make,
but the involved numbers are smaller by orders of magnitude. In the present example, this number\footnote{Despite
this number being not so small, we find a solution in less than 1 second on a 2.7 GHz Pentium computer. This means that one of the very first choices for the quarks allows for a hypercharge direction. Our experience shows that this is a general pattern.} is
\begin{equation}
{3 \choose 3} \times {12 \choose 6} \times {6 \choose 3} = 18,480.
\end{equation}
Denoting the directions corresponding
to the abelian factors by capital letters, our ansatz takes the form
\begin{equation}
Y = x_1 U_1 + x_2 U_2 + \ldots +  x_n U_n.
\label{eq:hypercharge_direction}
\end{equation}
The hypercharge of a particular representation is given by the scalar product of its highest weight
and $Y$. If we denote the highest weights of the left-handed quark doublets and of the right-handed quark singlets by $\Lambda_i$, $i=1,\ldots,9$, we obtain
\begin{equation}
\Lambda_i \cdot Y = x_1 \,\, \Lambda_i \cdot U_1 + x_2 \,\, \Lambda_i \cdot U_2 + \ldots +  x_n \,\,
\Lambda_i \cdot U_n,
\label{eq:quarks_corrent_hypercharge}
\end{equation}
which is a {\it system of linear diophantine equations}, provided we multiply the left- and right-hand-sides
by $3N$. {\it We will henceforth not explicitly mention that the preceding and all following equations are to be multiplied by $3N$.} The factor of $N$ was explained in a footnote on page \pageref{footnote:multiply_by_N}, and the factor of 3 is the smallest common multiple of the denominators of the hypercharge assignments. Now it becomes very clear that in taking these equations to be diophantine, we restrict our search to such exotics which have hypercharge values with a denominator not greater than 3. We will lift this constraint in Section \ref{sec:non-linear-constraints}.

\medskip

Although one may think that in our example, where we have 9 independent $\text{U}(1)$ directions, these 9 equations severely constrain the values of $x_i$, this is not true. In general, the system will be under-determined, and, as a matter of fact, in our case, is under-determined, since the quarks differ by localization, but not necessarily by the highest weights of their gauge representations.

\medskip

\item In order to account for the hypercharges of the leptons and Higgses, and for the absence of chiral
exotics, we set up necessary, but in general {\it not sufficient}, linear constraints:
\begin{equation}
\sum_{(\bs{3},\bs{2}),(\bsb{3},\bs{2})} Y = 1, \qquad \sum_{(\bs{3},\bs{1}),(\bsb{3},\bs{1})} Y = -2,
\qquad \sum_{(\bs{1},\bs{2})} Y = -3, \qquad \sum_{(\bs{1},\bs{1})} Y = 6.
\label{eq:linear_constraint}
\end{equation}
For readability, we use $Y$ in the above equations as a shorthand for $\Lambda_k \cdot Y$, where
$\Lambda_k$ runs over the highest weights of the representations in the sum.

Note that the sum over e.g.~the $(\bs{3},\bs{1})$ and $(\bsb{3},\bs{1})$ representations reduces to
that over the right-handed quarks $\uc$ and $\dc$ alone, since we assume pairs of exotic particles
to carry hypercharge assignments that are equal in magnitude but opposite in sign.

\medskip

\item Another constraint comes from the requirement that hypercharge be non-anomalous. If the model has
an anomalous $\text{U}(1)$ direction that we will denote by $\text{U}_{1A}$, we demand that it be orthogonal to $Y$:
\begin{equation}
\left(x_1 U_1 + x_2 U_2 + \ldots +  x_n U_n\right) \cdot \text{U}_{1A} = 0
\label{eq:lineq_for_U1A}
\end{equation}

\medskip

\item For solving the obtained system of linear diophantine equations, we developed {\tt C++} code
implementing the algorithms described in Ref.~\cite{Esmaeili::2001ab}.

\medskip

\item The solution to the system of linear diophantine equations is given by the sum of a particular
solution and a linear combination of (not necessarily linearly
independent, see Ref.~\cite{Esmaeili::2001ab}) homogeneous
solutions\footnote{For efficiency, we minimized the length of the
particular solution, and applied an LLL lattice basis
reduction \cite{LLLF82ab} on the homogeneous solutions using the NTL
libraries \cite{NTL06}.}, where the coefficients are all integers.
In most cases, there are no more than 2 homogeneous solutions, so
that we can simply iterate over the integer coefficients to obtain
candidates for hypercharge. Each candidate will surely fulfill
Eqs.~(\ref{eq:quarks_corrent_hypercharge}-\ref{eq:lineq_for_U1A}),
but may fail to give the correct hypercharge values on the leptons
and Higgses; furthermore, the spectrum may not be vector-like.

\medskip

\item Some remarks are in order. First, note that we explicitly make use of the coefficients being integers,
which is an assumption that is not fully justified. Second, we can of course not loop over all integer coefficients,
but only over a finite subset. The lattice basis reduction \cite{LLLF82ab} ensures that the integer coefficients are ``small.''
Nevertheless, by restricting ourselves to a finite set, we may miss an interesting solution. The general ansatz
presented in Section \ref{sec:non-linear-constraints} will not suffer from these restrictions.

\end{inparaenum}

\bigskip

A detailed discussion of our results is presented in Section \ref{sec:analyzing_the_models}. At this point,
let us briefly remark that we applied the strategy outlined above to the 1,767 three-generation\footnote{This number
is slightly higher than the 1,170 models listed in Ref.~\cite{Lebedev:2006kn}, since
we lifted the constraint that the Standard Model gauge symmetry be a subset of $\text{SO}(10)$. See Section \ref{sec:analyzing_the_models} for more details.}
models. For 1,114
models, we could establish the existence of a suitable hypercharge direction. Since a negative answer in the
case of the remaining 653 models does not prove that hypercharge does not exist, we will investigate these
models using the more sophisticated tools described in Section \ref{sec:non-linear-constraints}.

\subsection{The General Case with Non-linear Constraint Equations}
\label{sec:non-linear-constraints}

\begin{inparaenum}[(i)]

We now turn our attention to the most general ansatz for
constructing a suitable hypercharge direction for which we have a
vector-like spectrum.

\medskip

\item Again, we make an ansatz
\begin{equation}
Y = x_1 U_1 + x_2 U_2 + \ldots +  x_n U_n
\label{eq:hypercharge_direction_2}
\end{equation}
for the hypercharge direction in terms of the abelian group factors and demand that the linear constraints Eqs.~(\ref{eq:quarks_corrent_hypercharge}-\ref{eq:lineq_for_U1A}) be satisfied.

\medskip

\item The linear equations by no means exhaust the constraints we may require to be fulfilled by a vector-like spectrum. In particular, there are the cubic,
\begin{equation}
\sum_{(\bs{3},\bs{2}),(\bsb{3},\bs{2})} Y^3 = \frac{1}{9}, \qquad \sum_{(\bs{3},\bs{1}),(\bsb{3},\bs{1})} Y^3 = -\frac{56}{9}, \qquad \sum_{(\bs{1},\bs{2})} Y^3 = -3, \qquad \sum_{(\bs{1},\bs{1})} Y^3 = 24,
\label{eq:cubic_constraint}
\end{equation}
and the quintic,
\begin{equation}
\sum_{(\bs{3},\bs{2}),(\bsb{3},\bs{2})} Y^5 = \frac{1}{81}, \qquad \sum_{(\bs{3},\bs{1}),(\bsb{3},\bs{1})} Y^5 = -\frac{992}{81}, \qquad \sum_{(\bs{1},\bs{2})} Y^5 = -3, \qquad \sum_{(\bs{1},\bs{1})} Y^5 = 96,
\label{eq:quintic_constraint}
\end{equation}
constraints. As before, the sum over e.g.~all $(\bs{1},\bs{1})$
representations reduces to that of the right-handed electrons that
carry hypercharge +2, since the right-handed neutrinos have zero
hypercharge, and the exotic particles come in vector-like pairs so
that their contribution to the sum vanishes.

\medskip

\item It is important to note that in this approach we consider the linear and the non-linear equations on the same footing. In particular, we do not assume that the linear equations are diophantine. This fact adds to the calculational complexity of the problem, which scales with the number of variables.

\medskip

\item As in the case of the linear constraints, we scale $Y$ by $3N$, so that the coefficients in the polynomial equations are guaranteed to be integers. This is more calculational convenience than a conceptional necessity, since the computational tools we are using and that are to be described below have a far better performance for integer arithmetics.

\medskip

\item The fact that these equations are highly non-linear may at first seem discouraging. Luckily, there are efficient methods for solving systems of polynomial equations. The first step is the calculation of a Gr\"obner basis \cite{Buchberger:1965ab}, from which we can already read off the dimensionality of the solution set, and in particular, whether the system has a solution at all. To this end, we used the computer algebra system {\tt Singular} \cite{GPS05}. We present the details of the calculation in Appendix \ref{app:details_model_yes_1}.

\medskip

\item If there are no solutions to the constraint equations, this rigorously proves the absence of a $\text{U}(1)$ that may play the role of hypercharge. In the cases where solutions exist, we determine them numerically using Laguerre's algorithm as implemented by {\tt Singular} \cite{GPS05}. Eq.~(\ref{eq:hypercharge_direction}) then gives the corresponding hypercharge direction, for which we can explicitly check whether our criteria are satisfied or not.

\medskip

\item It turns out that in all cases, where the linear, cubic, and quintic equations were satisfied, the solution gives a vector-like spectrum. Demanding only the linear and cubic equations, however, does not guarantee that the spectrum is vector-like, as we have learned from experience in the course of our investigations.

\end{inparaenum}

\bigskip

In conclusion, let us remark that solving the non-linear constraint
equations adds 106 models to the list of 1,114 vector-like spectra
of three-generation Standard Models already constructed in Section
\ref{sec:linear-constraints}, so that the total number increases to
1,220. Also, this approach rigorously proves that for 547 out of
1,767 models, there is no $\text{U}(1)$ direction that can play the
role of hypercharge in a way that is compatible with low-energy
physics. In Section \ref{sec:results},
we discuss these results in greater detail.


\section{Phenomenology of the Models}
\label{sec:results}

\subsection{The Mini-Landscape Revisited}
\label{sec:minlandscape-revisited}

We consider a class of orbifold models with promising phenomenology \cite{Lebedev:2006kn}. In order to keep the discussion as clear as possible, we focus on one particular shift, namely $V^{\text{SO}(10),1}$, which incidentally gives the largest number of models.

\medskip

The results of the mini-landscape search \cite{Lebedev:2006kn} relevant to our present discussion can be summarized as follows. The shift $V^{\text{SO}(10),1}$ breaks
\begin{equation}
\text{E}_8 \times \text{E}_8' \ra \text{SO}(10)\times\text{SU}(2)^2\times \text{U}(1) \times \text{SO}(14)' \times\text{U}(1)'.
\label{eq:pattern_gauge_symmetry_breaking}
\end{equation}
If we allow for up to 2 Wilson lines, there are 21,841 inequivalent
models. What we mean by two models being inequivalent is explained in
Section \ref{sec:general_Y_search_in_minilandscape}.
\medskip

\begin{figure}[h]
\centering
\epsfig{figure=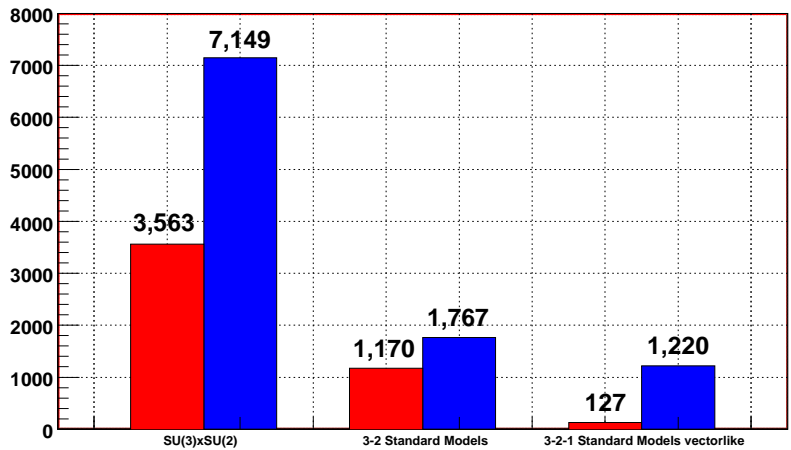, angle=0, scale=0.6}
\caption{Comparing the results of Ref.~\cite{Lebedev:2006kn} (red) to our findings (blue). Dropping the assumption of hypercharge coming from an $\text{SO}(10)$ leads to 10$\times$ more models.}
\label{fig:models_general_hypercharge}
\end{figure}

The gauge symmetries of 7,149 models contain $\text{SU}(3)\times\text{SU}(2)$ as a subgroup in at least one $\text{E}_8$, and thereof, 1,767 models have a three-generation spectrum. Since we do not know hypercharge yet, {\it three-generation model} in this context means that the spectrum, considering only color and weak isospin, has the right number of representations that can be identified with the particle content of the Standard Model. For a specific example of such a model, see e.g.~Tab.~\ref{tab:spectrum_model_yes_1} on page \pageref{tab:spectrum_model_yes_1}. We will refer to it as a 3-2 Standard Model.

\medskip

In the mini-landscape search \cite{Lebedev:2006kn}, our rationale had been to construct a
default candidate for hypercharge, namely that of Georgi-Glashow
$\text{SU}(5) \subset \text{SO}(10)$. To that end, we considered
3,563 models whose gauge symmetries contained
$\text{SU}(3)\times\text{SU}(2)$ {\it as a subgroup of}
$\text{SO}(10)$. Of this set, 1,170 models had a three-generation
spectrum w.r.t.~the $\text{SU}(3)\times\text{SU}(2) \subset
\text{SO}(10)$ gauge symmetry. Taking hypercharge to be that of the
Georgi-Glashow model, 127 models had a vector-like spectrum. Here
and in the following, we will call a three-generation model a 3-2-1 {\it
vector-like} Standard Model, if hypercharge has the correct values (cf.~Tab.~
\ref{tab:spectrum_SM}) on the quarks, leptons, and Higgses, and if
all the exotic states come in pairs such that the Standard Model
quantum numbers including hypercharge in each pair are equal in
magnitude but opposite in sign.

\subsection{The General Search for Hypercharge in the ``Fertile Patch''}
\label{sec:general_Y_search_in_minilandscape}

The starting point of our search is the set of 21,841 inequivalent
models. We call 2 models equivalent, if their spectra coincide. In this context, there are a couple of non-trivial issues that must be addressed. \begin{inparaenum}[(i)] \item The unbroken gauge group is typically the product of several simple factors. For each such factor, we have the choice whether to read its Dynkin diagram from left to right or from right to left, which corresponds to swapping all the representations with the complex conjugate ones. Thus, 2 spectra that differ only by complex conjugation w.r.t.~one or more gauge group factors are to be identified. \item Because of its highly symmetric Dynkin diagram, the algebra $\text{SO}(8)$ deserves special attention. Although the representations $\bs{8}_v$, $\bs{8}_s$, $\bs{8}_c$ are not equivalent, their tensor products are unaffected by a cyclic permutation, so we identify models whose spectra coincide after this permutation. \item In most cases, the same gauge group factor appears multiple times in the gauge symmetry. We identify those models whose spectra coincide when permuting the identical factors {\it in the same} $\text{E}_8$. \item Two models whose spectra are identical may differ in the localization of the particles. {\it At present, we do not distinguish between those models.} \item When checking the equivalence of models, we consider only the non-abelian charges. To what extent two models that coincide in all the non-abelian quantum numbers may differ w.r.t.~$\text{U}(1)$ charges is subject to further research.
\end{inparaenum}

\medskip

In the present case we want to be as general as possible, so, in contrast
to the mini-landscape search, we do not require that the Standard
Model gauge group be contained in $\text{SO}(10)$. There are 7,149
models whose unbroken gauge group contains
$\text{SU}(3)\times\text{SU}(2)$ as a subset of the first or
second $\text{E}_8$, and 1,767 of them are 3-2 Standard Models, see Fig.~\ref{fig:models_general_hypercharge}.

\medskip

For each of the 1,767 models, we first identify all representations of the
form $(\bs{3},\bs{2})$ and $(\bsb{3},\bs{1})$. Generically, the
number of these representations is greater than 3 and 6, respectively,
so we must make a choice as to what we call left-handed quark doublets
and right-handed up- and down-type anti-quark singlets. For each such
choice, we generate the linear, cubic, and quintic
constraint equations along the lines of Section
\ref{sec:hypercharge_general}. To get a flavor of the details of the
calculation, the reader is referred to the example in Appendix
\ref{app:details_model_yes_1}. Let us here briefly remark that
we are making the most general ansatz in that we treat the linear,
cubic, and quintic constraints on the same footing and that {\it we do not
assume} that the equations are diophantine.

\medskip

The following
cases need to be distinguished. \begin{inparaenum}[(i)] \item There
are only finitely many solutions. We pick any of the
solutions and calculate $Y$, see Eq.~(\ref{eq:hypercharge_direction_2}) on page \pageref{eq:hypercharge_direction_2}. \item The solutions are given by
continuous parameters, and the relations intertwining these parameters
are linear. We specialize to a numerical solution by
setting all independent parameters equal to zero. \item Again, the solutions are given by
continuous parameters, but this time, the relations intertwining the parameters
are non-linear. Since we cannot easily solve for the independent
parameters, we numerically find one special solution and calculate the
corresponding $Y$. Hypercharge directions which lead to irrational charges for the exotics are discarded.
\end{inparaenum}

\medskip

Since our goal is to establish the existence of a suitable
hypercharge direction, we stop at the first solution we find, i.e.~as
soon as the dimension of the ideal that describes the set of solutions to the system of
polynomial equations is greater than or equal to zero.

\medskip

Our findings are most unexpected. It turns out that 1,220 models
allow for a hypercharge direction for which the spectrum is
vector-like and contains three generations of quarks and leptons,
i.e.~we find ten times more models than in the mini-landscape case, see
Fig.~\ref{fig:models_general_hypercharge}. We were
unsuccessful in our search only for 547 models. In Section \ref{sec:analyzing_the_models},
we will analyze the 1,220 vector-like 3-2-1 Standard Models in greater detail.

\medskip

\subsection{Analyzing the Vector-like Models}
\label{sec:analyzing_the_models}

An important question is in what respect these
1,220 vector-like three-generation Standard Models differ from the
127 ones constructed previously, see Fig.~\ref{fig:models_general_hypercharge}. We will take a
closer look at some key properties of the 127 models that are
deemed desirable features for their phenomenology and see to what
extent they are realized in the larger set.

\subsubsection*{1,126 Models: Standard Model Gauge Group in $\bs{\text{SO}(10)}$}

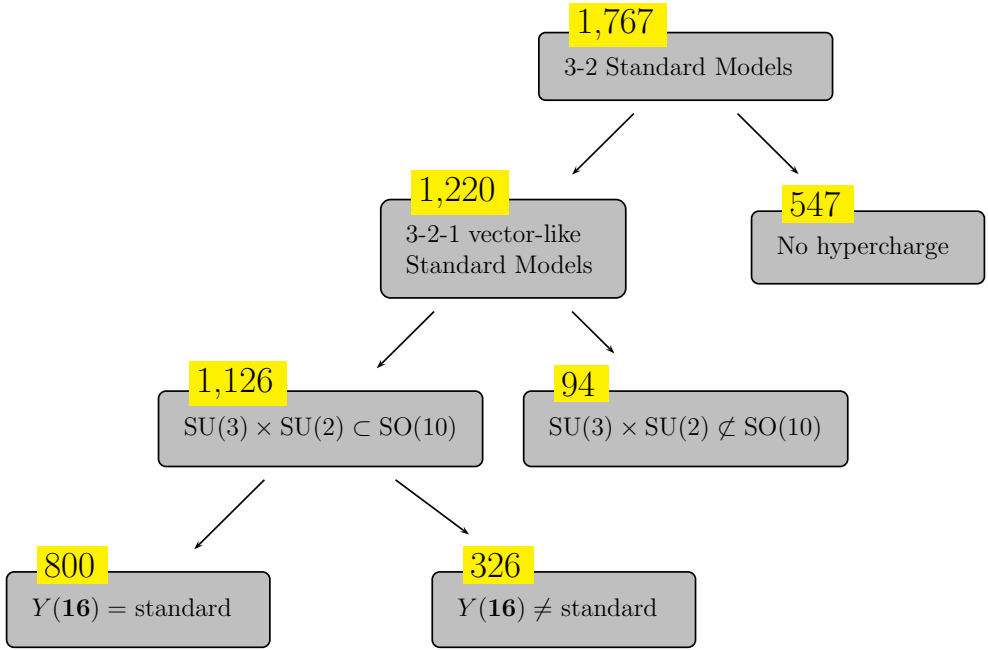
\begin{figure}[h!]
\hspace{1cm}\scalebox{0.8}{
\begin{pspicture}(-7,1)(5,-10)
\psset{unit=1cm}
\rput(+4,+0){\rnode{A}{\myBox{4cm}{\Large 1,767}{3-2 Standard Models}}}
\rput(+1,-3){\rnode{B}{\myBox{3.2cm}{\Large 1,220}{3-2-1 vector-like \\ Standard Models}}}
\rput(+7,-3){\rnode{C}{\myBox{3cm}{\Large 547}{No hypercharge}}}

\rput(-2,-6){\rnode{D}{\myBox{4.5cm}{\Large 1,126}{$\text{SU}(3)\times\text{SU}(2) \subset \text{SO}(10)$}}}
\rput(+4,-6){\rnode{E}{\myBox{4.5cm}{\Large 94}{$\text{SU}(3)\times\text{SU}(2) \not\subset \text{SO}(10)$}}}

\rput(-5,-9){\rnode{F}{\myBox{3.5cm}{\Large 800}{$Y(\bs{16}) =$ standard}}}
\rput(+2,-9){\rnode{G}{\myBox{3.5cm}{\Large 326}{$Y(\bs{16}) \not=$ standard}}}

\ncline[nodesepA=3pt, nodesepB=3pt]{->}{A}{B}
\ncline[nodesepA=3pt, nodesepB=10pt]{->}{A}{C}
\ncline[nodesepA=3pt, nodesepB=3pt]{->}{B}{D}
\ncline[nodesepA=3pt, nodesepB=10pt]{->}{B}{E}
\ncline[nodesepA=3pt, nodesepB=3pt]{->}{D}{F}
\ncline[nodesepA=3pt, nodesepB=10pt]{->}{D}{G}
\end{pspicture}}
\caption{Overview over the fraction of models passing our criteria.}
\label{fig:properties_vector-like_models}
\end{figure}

Remember that in Section \ref{sec:results} we had dropped the
assumption that the Standard Model gauge group be contained in
$\text{SO}(10)$. If we now reinstate this constraint, the number of
vector-like models decreases slightly from 1,220 to 1,126 (see
Fig.~\ref{fig:properties_vector-like_models}). Based on the number
of three-generation models in
Fig.~\ref{fig:models_general_hypercharge}, we could have expected to
see a more pronounced effect, i.e.~a more dramatic drop. This suggests that the
$\text{SO}(10)$ structure enhances the chances of a model to have
three generations.

\medskip

A word of caution is in order. Because of the structure of the gauge symmetry breaking (see Eq.~(\ref{eq:pattern_gauge_symmetry_breaking})), the $\text{SU}(3)$ factor of the Standard Model must necessarily be in $\text{SO}(10)$. As to the $\text{SU}(2)$ factor, it may lie outside of $\text{SO}(10)$, but can in many cases be related to one which lies in $\text{SO}(10)$ by an outer automorphism, see Appendix \ref{sec:hypercharge_outside_SO10} for a specific example. Thus, we may be tracing a feature that is mathematical in nature and bears no physical relevance.

\medskip

The same example in Appendix \ref{sec:hypercharge_outside_SO10}
illustrates a point which we had not anticipated.  Two
distinct sets of shifts and Wilson lines may lead to the same gauge
group and spectrum, but to different symmetry breaking patterns. The
consequences of this observation are rather bizarre. In the case of
the first model in Appendix \ref{sec:hypercharge_outside_SO10}, the
symmetry breaking pattern allows us to complete the Standard Model
gauge group to an intermediate $\text{SU}(5)$ and consequently to
derive Georgi-Glashow hypercharge, for which the spectrum is
vector-like and has three generations of quarks and leptons. As such,
it qualifies for our list of phenomenologically promising models. In
contrast, in the second model, there is no such intermediate
$\text{SU}(5)$ that we can use to construct a default hypercharge
direction\footnote{In the case of the mini-landscape study, this was
not a problem, since we {\it first} constructed Georgi-Glashow
hypercharge and {\it then} removed equivalent models.  As a result, we
ran over {\it all} models.}.

\medskip

In the early stages of our analysis, we had found only 1,094 models,
whose Standard Model gauge group was a subset of
$\text{SO}(10)$. After becoming aware of the subtleties described in
the previous paragraph, we took a closer look at the 126 models which,
loosely speaking, were not a subset of $\text{SO}(10)$. In 32 cases,
we could find a ``twin model'' such that the spectra coincide and the
gauge group is indeed a subset of $\text{SO}(10)$, thus giving us
1,126 models in total, see Fig.~\ref{fig:properties_vector-like_models}.

\subsubsection*{127 Models: Hypercharge Direction Purely due to $\bs{\text{SO}(10)}$}

Next we would like to know how the 127 vector-like three-generation
models of the mini-landscape \cite{Lebedev:2006kn} fit into this new picture.
Clearly, we expect them to be a subset of the 1,126 models, and this
is easily verified. By construction, these 127 models satisfy
\begin{equation}
Y_{\text{general}} - Y_\textsc{gg} \equiv 0,
\end{equation}
and thus give the standard hypercharges on the $\bs{16}$ of
$\text{SO}(10)$. We will now generalize this desirable feature.

\subsubsection*{800 Models: Not Purely $\bs{\text{SO}(10)}$, but Standard Charges on $\bs{16}$-plet}

After these preliminary considerations, we ask the question that is at the heart of the matter: Of the 1,126 models, in those cases where $Y_{\text{general}}$ is different from $Y_\textsc{gg}$, how does this difference affect the spectrum?

\medskip

Despite this difference not being zero, the presence of $\text{SO}(10)$ is too conspicuous to be ignored. We expect that the $\bs{16}$-plets in the first twisted sector will still give rise to 2 of the 3 families, and our best guess is that $Y_{\text{general}}$ is such that it coincides with $Y_\textsc{gg}$ on these representations.

\medskip

Indeed, calculating $Y_\textsc{gg}$ for the 1,126 models where we
have an intermediate $\text{SU}(5)$ and comparing it to
$Y_{\text{general}}$, we find that in 800 cases, it coincides with
$Y_\textsc{gg}$ when evaluated on the $\bs{16}$-plet:
\begin{equation}
Y_{\text{general}} - Y_\textsc{gg} \not\equiv 0, \qquad
(Y_{\text{general}} - Y_\textsc{gg}) (\bs{16}) \equiv 0
\end{equation}
 In
Appendix \ref{sec:difference_Y_general_and_Y_GG}, we present an
explicit example. The two hypercharge directions $Y_\textsc{gg}$ and
$Y_{\text{general}}$, given in Eq.~(\ref{eq:SO10_GG}) and
Eq.~(\ref{eq:general_Y_1_appB}), respectively, differ only in
entries in the second $\text{E}_8$. As such, they give the same
hypercharge on particles coming from the $\bs{16}$ of
$\text{SO}(10)$ that lies completely in the first $\text{E}_8$, but
do change the hypercharge on the other states. Comparing the two
spectra for the {\it same shift and Wilson lines, but different
choices for hypercharge,} namely either $Y_\textsc{gg}$ or
$Y_{\text{general}}$, we find that in the former case, the model not
only fails to be vector-like, but does not even contain
three generations, see Appendix
\ref{sec:difference_Y_general_and_Y_GG} for the details. On the
other hand, taking hypercharge to be $Y_{\text{general}}$, we get a
three-generation model that is vector-like.

\medskip

We expect that these 800 models will inherit most of the
properties of the 127 ones derived in Ref.~\cite{Lebedev:2006kn}. The question of gauge
coupling unification is the exception from the rule:
The extra components of $Y_{\text{general}}$ lying outside $\text{SO}(10)$
will have an impact on the value of the \textsc{gut} scale. We will address this
question in Section \ref{sec:gauge_coupling_unification}.

\subsubsection*{326 Models: Families do not Form Complete Multiplets of $\bs{\text{SO}(10)}$}

In the remaining 326 vector-like three-generation models, it is no
longer true that two families arise from $\bs{16}$'s of
$\text{SO}(10)$ in the first twisted sector, i.e.~these models are
structurally different from the 127 ones we considered so far as
paradigms.

\medskip

We looked at one example in detail. In summary, calculating the values of $Y_{\text{general}}$ on the $\bs{16}$-plet at the origin in the first twisted sector, we find that the states corresponding to $\uc$ and $\ec$ are missing. Perusing the four-dimensional spectrum, it turns out that the representation playing the role of the $\uc$ lives at a fixed point in the first twisted sector with no $\text{SO}(10)$ symmetry, whereas $\ec$ lives at the origin, but arises from a doublet representation outside of $\text{SO}(10)$.

\subsection{Gauge Coupling Unification}
\label{sec:gauge_coupling_unification}

The value of $\sin^2 \theta_w$ deserves some special attention,
because for one thing, it sets the \textsc{gut} scale in the theory, and, as a
consequence, it bears direct relevance to observable parameters at the electroweak scale.

\medskip

At the \textsc{gut} scale, the couplings constants $g$ and $g'$ are equal, if the generators of $\text{SU}(2)_L$ and $\text{U}(1)_Y$ are normalized in the same way. In our conventions, where
\begin{equation}
Q = T_{3L} + \frac{Y}{2},
\end{equation}
there is a constant of proportionality, $g = C g'$, which leads to
\begin{equation}
\sin^2 \theta_w = \frac{g'^2}{g^2 + g'^2} = \frac{1}{1+C^2}.
\end{equation}
The constant is given by the relative normalization of the generators,
\begin{equation}
\vectornorm{\frac{Y}{2}} = C \vectornorm{T_{3L}}.
\end{equation}
Up to a factor, the norm is given by the Killing form, see e.g.~\S13.1.2.~of Ref.~\cite{DiFrancesco:1997nk}. Since the vector $Y$ represents the sum over Cartan generators with the respective coefficients, the norm of the generator corresponding to $Y$ reduces to its vector norm. Evaluating the Killing form on the fundamental representation for $\text{SU}(2)_L$ canonically used in particle physics gives $\vectornorm{T_{3L}}^2 = 1/2$.

\medskip

\begin{figure}[h]
\centering
\epsfig{figure=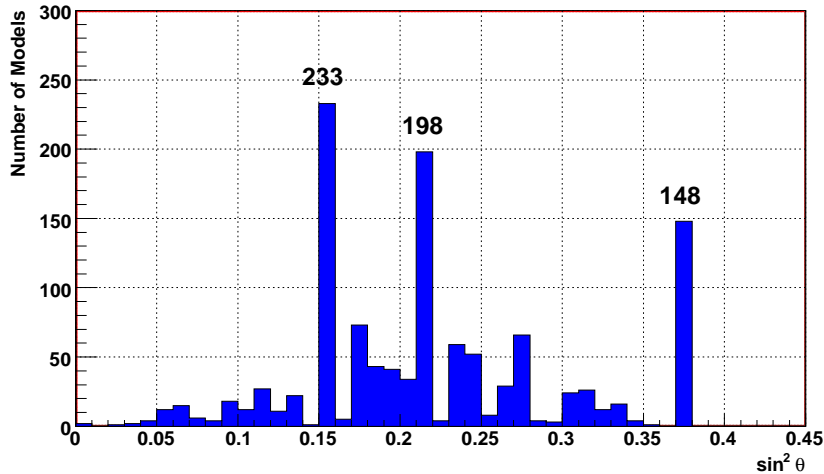, angle=0, scale=0.6}
\caption{The value of $\sin^2\theta_w$ at the \textsc{gut} scale for the 1,220 vector-like models.}
\label{fig:sin2theta}
\end{figure}

We calculated $\sin^2 \theta_w$ for the 1,220 models (see Fig.~\ref{fig:properties_vector-like_models}) for which a hypercharge direction exists. The results are presented in Fig.~\ref{fig:sin2theta}. For a judicious assessment of the data, the following remarks should be given some consideration.

\medskip

For a given model, there is typically more than one hypercharge
direction. In our general search in Section \ref{sec:analyzing_the_models}, we stopped at the first solution
that gave a vector-like 3-2-1 Standard Model. In most of the models, $\sin^2\theta_w$ is not only not equal to the standard \textsc{gut} value of $3/8$, but it is generically too small.

\medskip

Aiming at a higher value of $\sin^2 \theta_w$, our first ansatz has been to minimize the norm of $Y$. If there are only finitely many solutions to the constraint equations, we construct each hypercharge direction and pick the one with smallest norm. In many cases, however, the set of solutions is given by one or more parameters. In those cases where the parameters are given by linear relations, we use a simplex algorithm \cite{NAG} to find $Y$ with minimal norm. The algorithm requires a ``best guess'' as input and in some cases failed to find the minimal solution. There were also a couple of cases where the solutions were parameterized by non-linear equations, adding to the complexity of a numerical minimization.

\medskip

For this and other reasons, we decided to specifically search for those models which have a hypercharge direction such that $\sin^2 \theta_w = 3/8$. Bearing the results of Section \ref{sec:hypercharge_general} in mind, it is straightforward to implement this constraint (see Eq.~(\ref{eq:hypercharge_direction}) on page \pageref{eq:hypercharge_direction} for $Y$ in terms of the independent $\text{U}(1)$ directions):
\begin{equation}
\sin^2\theta_w = \frac{3}{8} \qquad \leftrightarrow \qquad \vectornorm{Y}^2 = \frac{10}{3}  \qquad \leftrightarrow \qquad \sum_{i,j} x_i x_j \,\, U_i\cdot U_j = \frac{10}{3}
\label{eq:minY}
\end{equation}

We simply add Eq.~(\ref{eq:minY}) to the set of constraint equations in Sections \ref{sec:linear-constraints} and \ref{sec:non-linear-constraints} and search for solutions. Naturally, we can restrict ourselves to the 1,220 vector-like Standard Models for which we know that hypercharge exists.

\medskip

The results are presented in Fig.~\ref{fig:sin2theta} on page \pageref{fig:sin2theta}. The values for $\sin^2 \theta_w$ in the histogram are ``biased'' in the sense that first, we specifically looked for the correct \textsc{gut} scale relation, and second, we picked the smallest value whenever the problem was amenable to numeric minimization. Had we listed the values without these priors, we would have more models with lower values of $\sin^2 \theta_w$.

\medskip

For 148 models, we find $\sin^2 \theta_w = 3/8$. This number is to
be compared to the 127 models of the mini-landscape (see
Fig.~\ref{fig:models_general_hypercharge} on page
\pageref{fig:models_general_hypercharge}), where hypercharge is
constructed from $\text{SO}(10)$ and the constraint on $\sin^2
\theta_w$ is thus satisfied by construction.  In the following we
take a closer look at the 21 extra models to assess in what respect,
if at all, they differ from the 127 ones.

\medskip

In 12 cases, hypercharge has components both in the first and in the second $\text{E}_8$. An explicit check shows that the
hypercharge assignments for the particles coming from the $\bs{16}$ of $\text{SO}(10)$ are non-standard in all cases.
This was to be expected, since the norm of hypercharge is the same as in the Georgi-Glashow case, but there exist
components outside of $\text{SO}(10)$. In 5 cases, $\text{SU}(3) \times \text{SU}(2) \subset \text{SO}(10)$. To see
how this fits into our present classification of the cases, see Fig.~\ref{fig:properties_vector-like_models} on
page \pageref{fig:properties_vector-like_models}.

\medskip

In 6 cases, hypercharge lives in the second $\text{E}_8$ and as such, the Standard Model is not a subset of the $\text{SO}(10)$
that is left over after the first symmetry breaking due to the shift alone. There is a fair chance that hypercharge may be in
some $\text{SO}(10)' \subset \text{SO}(14)' \subset \text{E}_8'$, see Eq.~(\ref{eq:pattern_gauge_symmetry_breaking}) on
page \pageref{eq:pattern_gauge_symmetry_breaking}, so we are following this lead. Indeed, in 5 cases the symmetry breaking pattern is
\begin{equation}
\text{E}_8' \quad \overset{V}{\longrightarrow} \quad \text{SO}(14)' \quad \overset{W_3}{\longrightarrow}
\quad \text{SU}(5)' \times \text{SU}(2)' \times \text{SU}(2)' \quad \overset{W_2}{\longrightarrow} \quad \text{SM} \times \text{hidden},
\end{equation}
and, moreover, hypercharge lies completely in $\text{SU}(5)'$, so we
are back to the Georgi-Glashow case.\footnote{Note, these differ
from the 127, since the Georgi-Glashow there was required to lie in
the first $\text{E}_8$.} In the one remaining case, the Standard
Model is a subgroup of $\text{SU}(4)' \times \text{SU}(4)'$.

\medskip

Most interestingly, in 3 cases hypercharge is all in the first $\text{E}_8$, but does not arise from Georgi-Glashow. The non-abelian gauge group factors of the Standard Model are not a subset of $\text{SO}(10)$, and in 2 cases the $\text{SU}(3)$ and $\text{SU}(2)$ factors are not only in different $\text{E}_8$'s, but the constructed hypercharge gives the standard values on the $\bs{16}$ of $\text{SO}(10)$. This observation is interesting enough to present the full details in Appendix \ref{app:details_weird_model}.


\section{Conclusions}

It has long been debated whether string theory is a theory of
everything or anything, and the present work tries to shed some
light on this important question. Admittedly, we have worked out the
details for a very special class of string theory models, namely
orbifold compactifications of the heterotic string. Despite all the
qualifications which will necessarily apply to our results, the big
advantage of our ansatz is that we have a complete picture in the
sense that we calculated all models in a well-defined framework and
based our conclusions on finite numbers, thereby avoiding any
discussion of a measure in the space of vacua.

\medskip

One result of our studies is that for 69\% of the 3-2 Standard
models, there exists a hypercharge direction such that the spectrum
becomes a 3-2-1 Standard Model and all exotics are vector-like. This
feature had hitherto been attributed to hypercharge arising from
Georgi-Glashow $\text{SU}(5)$ and has now been found to be a more
general result. About 66\% (or 800 of 1,220) of the 3-2-1 Standard
Models have the same characteristic features as those found in the
mini-landscape search \cite{Lebedev:2006kn} with the exception of
the Weinberg angle at the \textsc{gut} scale.

\medskip

The basic question we tried to answer in this publication can be summarized as follows: Given the Standard Model quantum
numbers and the absence of chiral exotics at low energies, do orbifold compactifications predict or at least prefer
values of $\sin^2\theta_w$ compatible with Grand Unification? In other words, in how many cases is the standard \textsc{gut}
value realized and to what extent is it correlated to the presence of an $\text{SO}(10)$ symmetry? Our results clearly
indicate that the most preferred value of $\sin^2\theta_w$ is in gross disagreement with Grand Unification and leads to
a \textsc{gut} scale which is too low. Nevertheless, a sizable fraction of the models attain the standard \textsc{gut}
value which lets us hope that there is some interesting physics yet to be explored.

\medskip

Realizing that it might have been too naive to expect string theory
to predict the Weinberg angle, we can add $\sin^2\theta_w = 3/8$ to
our list of priors and ask, given the Standard Model quantum
numbers, the absence of chiral exotics at low energies {\it and} the
Weinberg angle, do we necessarily have an $\text{SO}(10)$ as a local
symmetry? In this case, we find a strong correlation between
$\sin^2\theta_w = 3/8$ and an $\text{SO}(10)$ being realized at an
intermediate level in higher dimensions.  In fact, of the models
with the Standard Model spectrum, vector-like hypercharge and
$\sin^2\theta = 3/8$, 89\% have $\text{U}(1)_Y \subset \text{SU}(5)
\subset \text{SO}(10)$ in the first or second E$_8$. However, it
came somewhat as a surprise that there exist 16 models (11\%) which
do not quite fit into the usual \textsc{gut} picture and are
nevertheless consistent with unification. Moreover, there are 21
models in excess of the mini-landscape results \cite{Lebedev:2006kn}
which are a priori phenomenologically viable and should not be left
out of consideration in the quest for realistic models.

\medskip

Let us remark that the methods outlined in the present
publication have a broader range of application. These include the
search for (gauged) discrete family symmetries, for
Peccei-Quinn-type symmetries for the $\mu$-problem, and for a
suitable $B-L$ symmetry, a discrete subgroup of which can play the
role of $R$-parity. The latter case has already been worked out and
the results will be presented in a follow-up publication to the
mini-landscape project \cite{minilandscape-II}.

\medskip

The question of hypercharge normalization in the context of string theory has been addressed before, see e.g.~Ref.~\cite{Dienes:1995sq} and references therein. We are aware of the well-known discrepancy between unification at a string scale of order $10^{17}$ GeV and gauge coupling unification which occurs at $10^{16}$ GeV. There are, however, several reasons to believe that this discrepancy will be resolved by threshold corrections from Kaluza-Klein modes between a compactification scale and string scale, see e.g.~Refs.~\cite{Hall:2001pg,Contino:2001si,Dienes:1998vh,Kim:2002im} for orbifold \textsc{gut}s and Refs.~\cite{Hebecker:2004ce,Ibanez:1992hc,Kobayashi:2004ya,Kobayashi:2004ud} for string threshold corrections with asymmetrically large moduli. In either case the resolution comes from having Grand Unification in the extra dimensions and the Standard Model in 4 dimensions.

\medskip

Finally, a general search on a specific class of $\mathbb{Z}_3$ orbifolds has been performed in Ref.~\cite{Giedt:2001zw}. In this analysis, the author looks for Standard Model gauge structure with three families of quarks and leptons having the correct quantum numbers. In most cases, this requires a generalized hypercharge not in $\text{SU}(5)$ unlike what we find. However, it is difficult to compare our analysis to his, because he does not study those cases with $\sin^2\theta_w = 3/8$ in any detail, and in the general search, he does not distinguish between models with vector-like vs.~chiral exotics. In contrast, we first find the vector-like models with suitable hypercharge and then analyze $\sin^2\theta_w.$


\bigskip

\noindent {\bf Acknowledgments}
\noindent
We would like to thank James Gray and Nathan Salwen for useful discussions, and Angela Doerschlag for carefully reading the draft.
We are very much indebted to the Ohio Supercomputing Center for using their resources. This work was supported in part by the Department of
Energy under Grant No.\ DOE/ER/01545-873.


\newpage \clearpage \appendix


\section{Constructing Hypercharge for a Specific Model}
\label{app:details_model_yes_1}

We focus on one of the 21,841 models for shift $V^{\text{SO}(10),1}$
presented in Ref.~\cite{Lebedev:2006kn}, which we will henceforth
denote by $V_6$. Additionally, we have two Wilson lines, namely $W_3$ of order 3 and $W_2$ of order 2:

\begin{center}
\renewcommand{\arraystretch}{1.4}
\begin{tabular}{cccccccccccccccccc}
$V_6$     & = &  \big(    $\tfrac{2}{3}$ &   -$\tfrac{2}{3}$ &    $\tfrac{7}{6}$ &    $\tfrac{1}{6}$ &    $0$ &    $0$ &    $0$ &    $0$ &    $\tfrac{1}{3}$ &    $0$ &    $0$ &    $0$ &    $0$ &    $0$ &    $1$ &   -$1$  \big) \\
$W_3$     & = &  \big(   -$\tfrac{1}{6}$ &    $\tfrac{1}{6}$ &   -$\tfrac{1}{6}$ &   -$\tfrac{5}{6}$ &    $\tfrac{1}{6}$ &    $\tfrac{1}{6}$ &    $\tfrac{1}{6}$ &   -$\tfrac{1}{6}$ &    $\tfrac{2}{3}$ &    $0$ &   -$\tfrac{1}{3}$ &   -$\tfrac{1}{3}$ &   -$\tfrac{1}{3}$ &   -$\tfrac{1}{3}$ &   -$\tfrac{4}{3}$ &   -$\tfrac{2}{3}$  \big) \\
$W_2$     & = &  \big(   -$\tfrac{1}{2}$ &   -$1$ &    $1$ &    $0$ &    $\tfrac{1}{2}$ &   -$1$ &    $0$ &    $0$ &    $0$ &   -$\tfrac{1}{2}$ &    $0$ &    $0$ &    $0$ &    $0$ &    $0$ &   -$\tfrac{1}{2}$  \big) \\
\end{tabular}
\end{center}

The shift and Wilson lines break
\begin{equation}
\text{E}_8 \times \text{E}_8' \quad \longrightarrow \quad \text{SU}(3)\times\text{SU}(2)\times \text{U}(1)^5 \times \text{SU}(5)' \times {\text{U}(1)'}^4,
\end{equation}
and the spectrum is given by (omitting the $\text{U}(1)$ charges):
\begin{center}
\normalsize
\renewcommand{\arraystretch}{1.4}
\begin{tabular}{|r|r|r|r|}
\hline
$  3  \times (  \bs{3},  \bs{2},  \bs{1})$ &   $ 12  \times ( \bsb{3},  \bs{1},  \bs{1})$ &   $ 29  \times (  \bs{1},  \bs{2},  \bs{1})$ &   $ 8  \times (  \bs{1},  \bs{1}, \bs{5})$ \\
\hline
                                           &   $ 6  \times (  \bs{3},  \bs{1},  \bs{1})$ &    $136  \times (  \bs{1},  \bs{1},  \bs{1})$ &   $ 8  \times (  \bs{1},  \bs{1},  \bsb{5})$ \\
\hline
\end{tabular}
\end{center}

The simple roots of the unbroken gauge group are:
\begin{center}
\renewcommand{\arraystretch}{1.4}
\begin{tabular}{cccccccccccccccccc}
$\alpha_1$ & = &      \big( $0$ &    $0$ &    $0$ &    $0$ &    $0$ &    $0$ &    $0$ &    $0$ &    $0$ &    $0$ &    $0$ &    $0$ &    $0$ &    $1$ &   -$1$ &    $0$ \big) \\
$\alpha_2$ & = &      \big( $0$ &    $0$ &    $0$ &    $0$ &    $0$ &    $0$ &    $0$ &    $0$ &    $0$ &    $0$ &    $0$ &    $0$ &    $1$ &   -$1$ &    $0$ &    $0$ \big) \\
$\alpha_3$ & = &      \big( $0$ &    $0$ &    $0$ &    $0$ &    $0$ &    $0$ &    $0$ &    $0$ &    $0$ &    $0$ &    $0$ &    $1$ &   -$1$ &    $0$ &    $0$ &    $0$  \big)\\
$\alpha_4$ & = &      \big( $0$ &    $0$ &    $0$ &    $0$ &    $0$ &    $0$ &    $0$ &    $0$ &    $0$ &    $0$ &    $1$ &   -$1$ &    $0$ &    $0$ &    $0$ &    $0$  \big)\\
$\alpha_5$ & = &      \big( $0$ &    $0$ &    $0$ &    $0$ &    $0$ &    $0$ &    $1$ &    $1$ &    $0$ &    $0$ &    $0$ &    $0$ &    $0$ &    $0$ &    $0$ &    $0$ \big) \\
$\alpha_6$ & = &      \big( $0$ &    $0$ &    $0$ &    $0$ &    $0$ &    $1$ &   -$1$ &    $0$ &    $0$ &    $0$ &    $0$ &    $0$ &    $0$ &    $0$ &    $0$ &    $0$ \big) \\
$\alpha_7$ & = &      \big( $\tfrac{1}{2}$ &   -$\tfrac{1}{2}$ &   -$\tfrac{1}{2}$ &   -$\tfrac{1}{2}$ &   -$\tfrac{1}{2}$ &   -$\tfrac{1}{2}$ &   -$\tfrac{1}{2}$ &    $\tfrac{1}{2}$ &    $0$ &    $0$ &    $0$ &    $0$ &    $0$ &    $0$ &    $0$ &    $0$  \big)\\
\end{tabular}
\end{center}

The $\text{U}(1)$ directions are given by:
\begin{center}
\renewcommand{\arraystretch}{1.4}
\begin{tabular}{cccccccccccccccccc}
$U_1$ & = &     \big( $0$ &    $0$ &    $0$ &    $1$ &   -$1$ &    $0$ &    $0$ &    $0$ &    $0$ &    $0$ &    $0$ &    $0$ &    $0$ &    $0$ &    $0$ &    $0$  \big)\\
$U_2$ & = &     \big( $0$ &    $0$ &    $1$ &   -$1$ &    $0$ &    $0$ &    $0$ &    $0$ &    $0$ &    $0$ &    $0$ &    $0$ &    $0$ &    $0$ &    $0$ &    $0$  \big)\\
$U_3$ & = &     \big( $0$ &    $1$ &   -$1$ &    $0$ &    $0$ &    $0$ &    $0$ &    $0$ &    $0$ &    $0$ &    $0$ &    $0$ &    $0$ &    $0$ &    $0$ &    $0$  \big)\\
$U_4$ & = &     \big( $0$ &    $0$ &    $0$ &    $0$ &    $0$ &    $0$ &    $0$ &    $0$ &    $\tfrac{1}{2}$ &   -$\tfrac{1}{2}$ &   -$\tfrac{1}{2}$ &   -$\tfrac{1}{2}$ &   -$\tfrac{1}{2}$ &   -$\tfrac{1}{2}$ &   -$\tfrac{1}{2}$ &    $\tfrac{1}{2}$  \big)\\
$U_5$ & = &     \big( $0$ &    $0$ &    $0$ &    $0$ &    $0$ &    $0$ &    $0$ &    $0$ &    $0$ &    $0$ &    $0$ &    $0$ &    $0$ &    $0$ &    $0$ &   -$2$  \big)\\
$U_6$ & = &     \big( $0$ &    $0$ &    $0$ &    $0$ &    $0$ &    $0$ &    $0$ &    $0$ &    $0$ &    $1$ &    $0$ &    $0$ &    $0$ &    $0$ &    $0$ &   -$1$  \big)\\
$U_7$ & = &     \big( $\tfrac{1}{2}$ &   -$\tfrac{1}{2}$ &   -$\tfrac{1}{2}$ &   -$\tfrac{1}{2}$ &   -$\tfrac{5}{2}$ &    $\tfrac{3}{2}$ &    $\tfrac{3}{2}$ &   -$\tfrac{3}{2}$ &    $0$ &    $0$ &    $0$ &    $0$ &    $0$ &    $0$ &    $0$ &    $0$  \big)\\
$U_8$ & = &     \big( $\tfrac{1}{2}$ &   -$\tfrac{1}{2}$ &   -$\tfrac{1}{2}$ &   -$\tfrac{1}{2}$ &    $\tfrac{1}{2}$ &    $\tfrac{1}{2}$ &    $\tfrac{1}{2}$ &   -$\tfrac{1}{2}$ &    $0$ &    $0$ &    $0$ &    $0$ &    $0$ &    $0$ &    $0$ &    $0$  \big)\\
$U_9$ & = &     \big( $0$ &    $0$ &    $0$ &    $0$ &    $0$ &    $0$ &    $0$ &    $0$ &    $0$ &   -$2$ &    $1$ &    $1$ &    $1$ &    $1$ &    $1$ &   -$1$  \big)\\
\end{tabular}
\end{center}

\subsubsection*{Linear Diophantine Equations}

It is easy to check that the $\text{U}(1)$ directions are all
orthogonal to the simple roots, and furthermore belong to the
$\text{E}_8 \times \text{E}_8'$ lattice, thus satisfying criterion (\ref{it:i})
on page \pageref{it:i}. For one particular choice of
left- and right-handed quarks (which, due to space limitations, we cannot specify further, unless we list all the details of the spectrum, such as highest weights, localization, etc.), we obtain the following particular
\begin{center}
\renewcommand{\arraystretch}{1.4}
\begin{tabular}{cccccccccccccccccc}
$\tilde{U}_0$ & = &     \big(   -$\tfrac{3}{2}$ &    $\tfrac{3}{2}$ &    $\tfrac{3}{2}$ &    $\tfrac{3}{2}$ &    $\tfrac{3}{2}$ &   -$\tfrac{5}{2}$ &   -$\tfrac{5}{2}$ &    $\tfrac{5}{2}$ &    $0$ &   -$6$ &   -$1$ &   -$1$ &   -$1$ &   -$1$ &   -$1$ &    $1$  \big) \\
\end{tabular}
\end{center}
 and homogeneous solutions
\begin{center}
\renewcommand{\arraystretch}{1.4}
\begin{tabular}{cccccccccccccccccc}
$\tilde{U}_1$ & = &     \big(    $0$ &    $1$ &    $0$ &   -$1$ &    $0$ &    $0$ &    $0$ &    $0$ &   -$\tfrac{1}{2}$ &    $\tfrac{1}{2}$ &    $\tfrac{1}{2}$ &    $\tfrac{1}{2}$ &    $\tfrac{1}{2}$ &    $\tfrac{1}{2}$ &    $\tfrac{1}{2}$ &    $\tfrac{7}{2}$  \big) \\
 $\tilde{U}_2$ & = &     \big(   $0$ &   -$1$ &    $0$ &    $1$ &    $0$ &    $0$ &    $0$ &    $0$ &    $\tfrac{1}{2}$ &   -$\tfrac{1}{2}$ &    $\tfrac{1}{2}$ &    $\tfrac{1}{2}$ &    $\tfrac{1}{2}$ &    $\tfrac{1}{2}$ &    $\tfrac{1}{2}$ &    $\tfrac{3}{2}$  \big) \\
\end{tabular}
\end{center}
to the system of 14 equations corresponding to Eqs.~(\ref{eq:quarks_corrent_hypercharge}-\ref{eq:lineq_for_U1A}) in Section \ref{sec:linear-constraints}. Hypercharge is then given by
\begin{equation}
Y = \frac{1}{3} \, \big( \, \tilde{U}_0 + x_1 \, \tilde{U}_1 + x_2 \, \tilde{U}_2 \, \big)
\label{eq:app_general_sol_Y}
\end{equation}
for some $x_1, x_2 \in \mathbb{Z}$. We loop over all $-2 \leq x_1,
\, x_2 \leq 2$ and check whether $Y$ is such that it gives the
correct values on the quarks, leptons and Higgses, and whether the
exotics are vector-like. The first solution we find is
\begin{equation}
x_1 = 1, \qquad x_2 = 1,
\end{equation}
which corresponds to
\begin{equation}
\renewcommand{\arraystretch}{1.4}
\begin{tabular}{cccccccccccccccccc}
$Y$ & = &     \big(      -$\tfrac{1}{2}$  &  $\tfrac{1}{2}$  &  $\tfrac{1}{2}$   & $\tfrac{1}{2}$   & $\tfrac{1}{2}$  & -$\tfrac{5}{6}$  & -$\tfrac{5}{6}$  &  $\tfrac{5}{6}$   & $0$  & -$2$  &  $0$  &  $0$   & $0$   & $0$  &  $0$ &   $2$ \big). \\
\end{tabular}
\label{eq:general_Y_1}
\end{equation}
The spectrum, now including hypercharge, reads
\begin{center}
\normalsize
\renewcommand{\arraystretch}{1.4}
\begin{tabular}{|r|r|r|r|r|r|}
\hline
$  3  \times (  \bs{3},  \bs{2})_{1/3}$ &   $ 5  \times ( \bsb{3},  \bs{1})_{\text{-}4/3}$ &   $ 7  \times ( \bsb{3},  \bs{1})_{\phantom{\text{-}}2/3}$         &   $ 16  \times (  \bs{1}, \bs{2})_{\text{-}1}$ &   $ 45  \times (  \bs{1},  \bs{1})_{\phantom{\text{-}}2}$ &   $ 129  \times (  \bs{1},  \bs{1})_0$ \\
\hline
                                        &   $ 2  \times (  \bs{3},  \bs{1})_{\phantom{\text{-}}4/3}$ &           $ 4  \times (  \bs{3},  \bs{1})_{\text{-}2/3}$ &   $13  \times (  \bs{1},  \bs{2})_{\phantom{\text{-}}1}$ &   $ 42  \times (  \bs{1},  \bs{1})_{\text{-}2}$  &   \\
\hline
\end{tabular}
\end{center}
where we have counted the representations of the hidden gauge group as multiplicities for the Standard Model particles.
\bigskip

Several remarks are in order. First, let us note that, had we not
minimized the length of the particular solution and applied an LLL
reduction \cite{LLLF82ab} to the homogeneous solutions, we would have had to cover a considerably larger set of integers for $x_1$ and $x_2$. >From experience we know that the loss of time in this case can be considerable.

\medskip

Second, the presented model is such that the Standard Model gauge
group {\it is} a subset of $\text{SO}(10)$, but that {\it none} of
the $\text{U}(1)$ directions stemming from the \textsc{gut} gives a
three-generation model with vector-like exotics. Under this aspect,
the construction of a general hypercharge direction may give useful phenomenological insights.

\subsubsection*{Linear Diophantine Equations and the Cubic Constraints}

Despite the fact that we were able to construct a suitable hypercharge
direction using only the linear constraints, this is not true in
general. To demonstrate the use of the non-linear constraints, we
pursue the present example further. Starting from the general solution
Eq.~(\ref{eq:app_general_sol_Y}) and substituting it into the cubic
constraint Eq.~(\ref{eq:cubic_constraint}) on page \pageref{eq:cubic_constraint}, we obtain a system of 4 polynomial equations\footnote{To generate and manipulate polynomial equations, we have found {\tt GiNaC} \cite{Bauer:2000cp} to be a very useful tool.} in 2 variables, which are too complicated even to attempt to write them down. However, calculating a Gr\"obner basis\footnote{Giving an introduction to Gr\"obner bases is beyond the scope of the present publication. For a general introduction, see Refs.~\cite{Adams:1996groebner, Becker:1993groebner}. We would like to point out Ref.~\cite{Gray:2006gn} that made us aware of Gr\"obner bases and their applications in physics.} yields the elegantly short and simple form
\begin{equation}
x_1^2 - 2 x_1 x_2 + x_2^2 = 0,
\end{equation}
from which we see that the system has a one-dimensional set of
solutions, e.g.~$x_1 = x_2 = 1$, as found before. One should not be
misled to think that satisfying the linear and cubic constraints
is sufficient for the spectrum to be three-generation and vector-like.
In the course of our work, we encountered cases in which the {\it
constraints were fulfilled}, but nevertheless the {\it spectrum
failed to be three-generation or vector-like}.

\subsubsection*{Linear Diophantine Equations, the Cubic and the Quintic Constraints}

Finally, requiring that the quintic constraints Eq.~(\ref{eq:quintic_constraint}) be satisfied leaves us with only a finite number of solutions. Again, the constraints are too long to list, but after computing the Gr\"obner basis, we obtain
\begin{gather}
\scriptstyle
x_1^2-2 x_1 x_2+x_2^2 = 0,\\
\scriptstyle
70 x_1 x_2^4-45 x_2^5+8 x_1 x_2^3-13 x_2^4-156 x_1 x_2^2+70 x_2^3+8 x_1 x_2+30 x_2^2+70 x_1+7 x_2-49 = 0,\\
\scriptstyle
875 x_2^6-450 x_2^5+2880 x_1 x_2^3-5835 x_2^4-1728 x_1 x_2^2+4004 x_2^3-5184 x_1 x_2+7461 x_2^2+4032 x_1-6594 x_2+539 = 0.
\end{gather}
The only solutions to this system are
\begin{equation}
x_1 = x_2 = 1 \qquad \text{and} \qquad x_1 = x_2 = -\dfrac{7}{5}.
\end{equation}
The first solution is the one we already know, and calculating the hypercharge direction corresponding to the second solution, we obtain
\begin{equation}
\renewcommand{\arraystretch}{1.4}
\begin{tabular}{cccccccccccccccccc}
$Y$ & = & \big( -$\tfrac{1}{2}$ &    $\tfrac{1}{2}$ &    $\tfrac{1}{2}$ &    $\tfrac{1}{2}$ &    $\tfrac{1}{2}$ &   -$\tfrac{5}{6}$ &   -$\tfrac{5}{6}$ &    $\tfrac{5}{6}$ &    $0$ &   -$2$ &   -$\tfrac{4}{5}$ &   -$\tfrac{4}{5}$ &   -$\tfrac{4}{5}$ &   -$\tfrac{4}{5}$ &   -$\tfrac{4}{5}$ &   -$2$ \big)
\end{tabular}
\label{eq:general_Y_2}
\end{equation}
leading to the spectrum
\begin{center}
\normalsize
\renewcommand{\arraystretch}{1.4}
\begin{tabular}{|r|r|r|r|r|}
\hline
$  3  \times (  \bs{3},  \bs{2})_{1/3}$ &   $ 5  \times ( \bsb{3},  \bs{1})_{\text{-}4/3}$ &    $ 16  \times (  \bs{1}, \bs{2})_{\text{-}1}$ &   $ 30  \times (  \bs{1},  \bs{1})_{\phantom{\text{-}}2}$ &   $ 15  \times (  \bs{1},  \bs{1})_{\phantom{\text{-}}6/5}$\\
\hline
                                        &   $ 2  \times (  \bs{3},  \bs{1})_{\phantom{\text{-}}4/3}$ &             $13  \times (  \bs{1},  \bs{2})_{\phantom{\text{-}}1}$ &   $ 27  \times (  \bs{1},  \bs{1})_{\text{-}2}$  &   $ 15  \times (  \bs{1},  \bs{1})_{\text{-}6/5}$\\
\hline
                                        &   $ 7  \times (  \bsb{3},  \bs{1})_{\phantom{\text{-}}2/3}$ &    &   $ 79  \times (  \bs{1},  \bs{1})_{\phantom{\text{-}}0}$  &   $ 25  \times (  \bs{1},  \bs{1})_{\phantom{\text{-}}4/5}$\\
\hline
                                        &   $ 4  \times (  \bs{3},  \bs{1})_{\text{-}2/3}$ &               &                                                   &   $ 25  \times (  \bs{1},  \bs{1})_{\text{-}4/5}$\\
\hline
\end{tabular}
\end{center}
which is a three-generation model with vector-like exotics. Thus, we
see that our assumption that the coefficients in
Eq.~(\ref{eq:app_general_sol_Y}) be integers is not fully justified.

\subsubsection*{The Linear, Cubic and Quintic Constraints on the Same Footing}

In light of these findings, we review our assumptions entering the foregoing calculation. Choosing the $\text{U}(1)$ directions to be root vectors of $\text{E}_8 \times \text{E}_8'$ is safe, since we can always find $n$ such directions such that together with the roots of the unbroken gauge group they span the full 16-dimensional lattice. For the linear constraints, we will obtain a system of equations with {\it integer coefficients}, but we must now drop the assumption that the solutions are given by integers, i.e.~the system is {\it not necessarily diophantine.} For the cubic and quintic constraints, we had not used assumed the solutions to be integers at all; therefore, are no modifications.

\medskip

In practice, this means that we cannot a priori reduce the system from 9 to 2 variables by solving the diophantine equations, but must rather consider the linear constraints on the same footing as the cubic and quintic ones. Starting with the general ansatz Eq.~(\ref{eq:hypercharge_direction}), we thus obtain a system of 12 polynomial equations in 9 variables. Pursuing the same calculations as previously outlined, one can show that there are exactly 2 solutions, corresponding to
\begin{equation}
x_1 = x_2 = x_3 = x_4 = x_5 = x_9 = 0, \quad x_6 = -6, \quad x_7 = -1, \quad x_8 = -2,
\end{equation}
and
\begin{equation}
x_1 = x_2 = x_3 = x_4 = 0, \quad x_5 = \dfrac{48}{5}, \quad x_6 = -\dfrac{54}{5}, \quad x_7 = -1, \quad x_8 = -2, \quad x_9 = -\dfrac{12}{5}.
\end{equation}
Using the nine $\text{U}(1)$ directions listed at the beginning of this section, we recover exactly the 2 hypercharge directions which we have already constructed, thereby proving that these are the most general solutions.


\newpage

\section{Where did We Lose the Interesting Models?}
\label{sec:difference_Y_general_and_Y_GG}

Consider the model given at the beginning of Appendix
\ref{app:details_model_yes_1}. We will now address the question to
what extent the general hypercharge directions given in
Eqs.~(\ref{eq:general_Y_1},\ref{eq:general_Y_2}) are connected to
the presence of an underlying $\text{SO}(10)$ symmetry. More
specifically, we will explain why this model was not found by the
analysis presented in Ref.~\cite{Lebedev:2006kn}, despite the fact
that its Standard Model gauge group is contained in $\text{SO}(10)$
and there exists a hypercharge direction such that its spectrum is
vector-like (see Appendix \ref{app:details_model_yes_1}).

\medskip

For the following discussion, only the first $\text{E}_8$ factor is relevant. The shift $V_6$ breaks
\begin{equation}
\text{E}_8 \,\,\overset{V_6}{\longrightarrow}\,\, \text{SO}(10) \times \text{SU}(2) \times \text{SU}(2) \times \text{U}(1),
\end{equation}
or in terms of Dynkin diagrams:

\begin{center}
\unitlength=.8mm
\begin{picture}(107,21)
\thicklines
\put(2, 5){\circle{4}} \put(17,5){\circle{4}} \put(32,5){\circle{4}} \put(32,20){\circle{4}} \put(47,5){\circle{4}} \put(62,5){\circle{4}} \put(77, 5){\circle{4}} \put(92, 5){\circle{4}} \put(107, 5){\circle{4}}
\put(4,5){\line(1,0){11}} \put(19,5){\line(1,0){11}} \put(32,7){\line(0,1){11}} \put(34,5){\line(1,0){11}} \put(49,5){\line(1,0){11}} \put(64,5){\line(1,0){11}} \put(79,5){\line(1,0){11}} \put(94,5){\dashline{2}(0,0)(11,0)}
\put(0,-1){$\alpha_1$} \put(15,-1){$\alpha_2$} \put(30,-1){$\alpha_3$} \put(35,19){$\alpha_8$} \put(45,-1){$\alpha_4$} \put(60,-1){$\alpha_5$} \put(75,-1){$\alpha_6$} \put(90,-1){$\alpha_7$} \put(105,-1){$\alpha_0$}
\put(59,2){\line(1,1){6}} \put(59,8){\line(1,-1){6}}
\put(89,2){\line(1,1){6}} \put(89,8){\line(1,-1){6}}
\end{picture}
\end{center}

The Wilson lines $W_3$ and $W_2$ break \text{SO}(10) down to the Standard Model:
\begin{equation}
\text{SO}(10) \,\,\overset{W_3}{\longrightarrow}\,\, \text{SU}(4) \times \text{SU}(2) \times \text{U}(1) \,\,\overset{W_2}{\longrightarrow}\,\, \text{SU}(3) \times \text{SU}(2) \times \text{U}(1)^2
\label{eq:SO10_to_SM}
\end{equation}

Again, this is most conveniently visualized in terms of the associated Dynkin diagram:

\begin{center}
\unitlength=.8mm
\begin{picture}(47,21)
\thicklines
\put(2, 5){\circle{4}} \put(17,5){\circle{4}} \put(32,5){\circle{4}} \put(32,20){\circle{4}} \put(47,5){\circle{4}}
\put(4,5){\line(1,0){11}} \put(19,5){\line(1,0){11}} \put(32,7){\line(0,1){11}} \put(34,5){\line(1,0){11}}
\put(0,-1){$\alpha_1$} \put(15,-1){$\alpha_2$} \put(30,-1){$\alpha_3$} \put(35,19){$\alpha_8$} \put(45,-1){$\alpha_4$}
\put(14,2){\line(1,1){6}} \put(14,8){\line(1,-1){6}}
\put(44,2){\line(1,1){6}} \put(44,8){\line(1,-1){6}}
\put(15,10){$W_3$}
\put(45,10){$W_2$}
\end{picture}
\end{center}

This clearly shows that the Standard Model gauge group {\it is
indeed a subset of} $\text{SO}(10)$. Nevertheless, perusing the list
of 127 vector-like models given in Ref.~\cite{Lebedev:2006kn}, we
find that this model is missing.

\medskip

To see why we missed this model, we adopt the following strategy. First, we construct the standard $\text{SO}(10)$ hypercharge as we did in Ref.~\cite{Lebedev:2006kn}. Then, we compare this hypercharge to the general one, given in Appendix \ref{app:details_model_yes_1}, Eqs.~(\ref{eq:general_Y_1},\ref{eq:general_Y_2}).

\medskip

Standard $\text{SO}(10)$ hypercharge is a linear combination of the Cartan generators that lies in $\text{SU}(5)$ and commutes with the $\text{SU}(3) \times \text{SU}(2)$ step operators. To find this hypercharge direction, we complete the Standard Model in all possible ways to an $\text{SU}(5)$ symmetry. Up to a multiplicative factor, hypercharge will then be given by the dual of the root which completes the diagram. Having $\text{SO}(10)$ in mind, we have a default choice:

\begin{center}
\begin{tabular}{ccccc}
{
\unitlength=.8mm
\begin{picture}(47,21)
\thicklines
\put(2, 5){\circle{4}} \put(32,5){\circle{4}} \put(32,20){\circle{4}}
\put(32,7){\line(0,1){11}}
\put(0,-1){$\gamma$} \put(30,-1){$\beta_2$} \put(35,19){$\beta_1$}
\end{picture}
} &  \raisebox{4ex}[-3ex]{ $\bs{\longrightarrow}$ } & {
\unitlength=.8mm
\begin{picture}(47,21)
\thicklines
\put(2, 5){\circle{4}} \put(17,5){\circle{4}} \put(32,5){\circle{4}} \put(32,20){\circle{4}}
\put(4,5){\dashline{2}(0,0)(11,0)} \put(19,5){\dashline{2}(0,0)(11,0)} \put(32,7){\line(0,1){11}}
\put(0,-1){$\gamma$} \put(15,-1){$\alpha$} \put(30,-1){$\beta_2$} \put(35,19){$\beta_1$}
\end{picture}
} &  \raisebox{4ex}[-3ex]{ $\bs{\longrightarrow}$ } & \raisebox{2ex}[-3ex]{ {
\unitlength=.8mm
\begin{picture}(47,21)
\thicklines
\put(2, 5){\circle{4}} \put(17,5){\circle{4}} \put(32,5){\circle{4}} \put(47,5){\circle{4}}
\put(4,5){\dashline{2}(0,0)(11,0)} \put(19,5){\dashline{2}(0,0)(11,0)} \put(34,5){\line(1,0){11}}
\put(0,-1){$\gamma$} \put(15,-1){$\alpha$} \put(30,-1){$\beta_2$} \put(45,-1){$\beta_1$}
\end{picture}
} }
\end{tabular}
\end{center}

The hypercharge direction is then given by
\begin{equation}
\text{U}(1)_{\text{SO}(10)} = \frac{5}{3} \alpha^* = \sum_{j} \frac{5}{3} \, (A_{\text{SU}(5)}^{-1})_{2j} \, \alpha_j = \frac{1}{3} \left( 3\gamma + 6\alpha + 4\beta_1 + 2\beta_2 \right),
\label{eq:U1GG}
\end{equation}
or if we substitute for the $\text{SU}(5)$ roots,
\begin{equation}
\renewcommand{\arraystretch}{1.4}
\begin{tabular}{cccccccccccccccccc}
$Y_{\textsc{gg}}$ & = & \big( -$\tfrac{1}{2}$ &    $\tfrac{1}{2}$ &    $\tfrac{1}{2}$ &    $\tfrac{1}{2}$ &    $\tfrac{1}{2}$ &   -$\tfrac{5}{6}$ &   -$\tfrac{5}{6}$ &    $\tfrac{5}{6}$ &    $0$ &   $0$ &   $0$ &   $0$ &   $0$ &   $0$ &   $0$ &   $0$ \big).
\end{tabular}
\label{eq:SO10_GG}
\end{equation}

Compare this result to the hypercharge direction found in Eq.~(\ref{eq:general_Y_1}) in Appendix \ref{app:details_model_yes_1} as the result of a general search:
\begin{equation}
\renewcommand{\arraystretch}{1.4}
\begin{tabular}{cccccccccccccccccc}
$Y_{\text{general}}$ & = &     \big(      -$\tfrac{1}{2}$  &  $\tfrac{1}{2}$  &  $\tfrac{1}{2}$   & $\tfrac{1}{2}$   & $\tfrac{1}{2}$  & -$\tfrac{5}{6}$  & -$\tfrac{5}{6}$  &  $\tfrac{5}{6}$   & $0$  & -$2$  &  $0$  &  $0$   & $0$   & $0$  &  $0$ &   $2$ \big) \\
\end{tabular}
\label{eq:general_Y_1_appB}
\end{equation}

Remarkably, the two hypercharge directions differ only in entries in the second $\text{E}_8$. To assess what this means for the phenomenology of the model at hand, we recalculate the spectra, where in one case, we have chosen hypercharge to be $Y_{\textsc{gg}}$,
\begin{center}
\normalsize
\renewcommand{\arraystretch}{1.4}
\begin{tabular}{|r|r|r|r|r|r|}
\hline
$  3  \times (  \bs{3},  \bs{2})_{1/3}$ &   $ 4  \times ( \bsb{3},  \bs{1})_{\text{-}4/3}$ &   $ 3  \times ( \bsb{3},  \bs{1})_{\phantom{\text{-}}2/3}$         &   $ 9  \times (  \bs{1}, \bs{2})_{\text{-}1\phantom{/3}}$ &   $ 2  \times (  \bs{1},  \bs{1})_{\phantom{\text{-}}2\phantom{/3}}$ &   $ 103  \times (  \bs{1},  \bs{1})_0$ \\
\hline
                                        &   $ 2  \times (  \bs{3},  \bs{1})_{\text{-}2/3}$ &           $ 4  \times (  \bs{3},  \bs{1})_{\phantom{\text{-}}0\phantom{/1}}$ &   $5  \times (  \bs{1},  \bs{2})_{\phantom{\text{-}}1\phantom{/3}}$ &   $ 6  \times (  \bs{1},  \bs{1})_{\phantom{\text{-}}4/3}$  &   \\
\cline{2-5}
                                        &                                                  &           $ 5  \times (  \bsb{3},  \bs{1})_{\phantom{\text{-}}0\phantom{/3}}$ &   $10  \times (  \bs{1},  \bs{2})_{\text{-}1/3}$ &   $ 6  \times (  \bs{1},  \bs{1})_{\text{-}4/3}$  &   \\
\cline{3-5}
                                        &                                                  &                                                        &   $5  \times (  \bs{1},  \bs{2})_{\phantom{\text{-}}1/3}$ &   $ 37  \times (  \bs{1},  \bs{1})_{\phantom{\text{-}}2/3}$  &   \\
\cline{4-5}
                                        &                                                  &                                                        &                                                  &   $ 62  \times (  \bs{1},  \bs{1})_{\text{-}2/3}$  &   \\
\hline
\end{tabular}
\end{center}

and in the other case, we have chosen hypercharge to be $Y_{\text{general}}$:
\begin{center}
\normalsize
\renewcommand{\arraystretch}{1.4}
\begin{tabular}{|r|r|r|r|r|r|}
\hline
$  3  \times (  \bs{3},  \bs{2})_{1/3}$ &   $ 5  \times ( \bsb{3},  \bs{1})_{\text{-}4/3}$ &   $ 7  \times ( \bsb{3},  \bs{1})_{\phantom{\text{-}}2/3}$         &   $ 16  \times (  \bs{1}, \bs{2})_{\text{-}1}$ &   $ 45  \times (  \bs{1},  \bs{1})_{\phantom{\text{-}}2}$ &   $ 129  \times (  \bs{1},  \bs{1})_0$ \\
\hline
                                        &   $ 2  \times (  \bs{3},  \bs{1})_{\phantom{\text{-}}4/3}$ &           $ 4  \times (  \bs{3},  \bs{1})_{\text{-}2/3}$ &   $13  \times (  \bs{1},  \bs{2})_{\phantom{\text{-}}1}$ &   $ 42  \times (  \bs{1},  \bs{1})_{\text{-}2}$  &   \\
\hline
\end{tabular}
\end{center}

In the case where hypercharge is purely $\text{SO}(10)$, the
spectrum fails to contain three generations of quarks and leptons
and is not vector-like either.

\medskip

Looking at the top row of the first spectrum, it catches one's eye that the particle content corresponds to that of 2 complete 16-plets of $\text{SO}(10)$, so our best guess is that 2 families still come from 16-plets localized at the first twisted sector and that the extra component of $Y_{\text{general}}$ does not affect these states at all. Indeed, this is easily verified by switching off the Wilson lines and evaluating $Y_{\text{general}} - Y_{\textsc{gg}}$ on these 16-plets, which gives zero. The same difference is non-zero on more than half of the other states. In particular, it is non-zero on the right-handed electron of the third family, as a comparison of the two spectra shows. As for the remaining states of the third family, there exist Standard Model representations with the right quantum numbers on which this difference vanishes.

\subsubsection*{Flipped $\bs{\text{SU}(5)}$}

Consider the discussion preceding Eq.~(\ref{eq:U1GG}). There is a second, not-so-obvious choice to complete the Standard Model Dynkin diagram to an $\text{SU}(5)$ symmetry and subsequently to obtain a possible hypercharge direction:

\begin{center}
\begin{tabular}{ccccc}
{
\unitlength=.8mm
\begin{picture}(47,21)
\thicklines
\put(2, 5){\circle{4}} \put(32,5){\circle{4}} \put(32,20){\circle{4}}
\put(32,7){\line(0,1){11}}
\put(0,-1){$\gamma$} \put(30,-1){$\beta_2$} \put(35,19){$\beta_1$}
\end{picture}
} &  \raisebox{4ex}[-3ex]{ $\bs{\longrightarrow}$ } & {
\unitlength=.8mm
\begin{picture}(47,21)
\thicklines
\put(2, 5){\circle{4}} \put(2,20){\circle{4}} \put(32,5){\circle{4}} \put(32,20){\circle{4}}
\put(2,7){\dashline{2}(0,0)(0,11)} \put(4,20){\dashline{2}(0,0)(26,0)} \put(32,7){\line(0,1){11}}
\put(0,-1){$\gamma$} \put(-6,19){$\alpha$} \put(30,-1){$\beta_2$} \put(35,19){$\beta_1$}
\end{picture}
} &  \raisebox{4ex}[-3ex]{ $\bs{\longrightarrow}$ } & \raisebox{2ex}[-3ex]{ {
\unitlength=.8mm
\begin{picture}(47,21)
\thicklines
\put(2, 5){\circle{4}} \put(17,5){\circle{4}} \put(32,5){\circle{4}} \put(47,5){\circle{4}}
\put(4,5){\dashline{2}(0,0)(11,0)} \put(19,5){\dashline{2}(0,0)(11,0)} \put(34,5){\line(1,0){11}}
\put(0,-1){$\gamma$} \put(15,-1){$\alpha$} \put(30,-1){$\beta_1$} \put(45,-1){$\beta_2$}
\end{picture}
} }
\end{tabular}
\end{center}

After a short calculation analogous to the one following Eq.~(\ref{eq:U1GG}), we obtain:
\begin{equation}
\renewcommand{\arraystretch}{1.4}
\begin{tabular}{cccccccccccccccccc}
$Y_{\text{flipped}}$ & = & \big( $\tfrac{1}{2}$ &    -$\tfrac{1}{2}$ &    -$\tfrac{1}{2}$ &    -$\tfrac{1}{2}$ &    $\tfrac{3}{2}$ &   $\tfrac{1}{6}$ &   $\tfrac{1}{6}$ &    -$\tfrac{1}{6}$ &    $0$ &   $0$ &   $0$ &   $0$ &   $0$ &   $0$ &   $0$ &   $0$ \big)
\end{tabular}
\label{eq:SU5_flipped}
\end{equation}

This hypercharge direction corresponds to flipped $\text{SU}(5)$, as can easily be shown by evaluating it on the $\bs{16}$ of $\text{SO}(10)$. In comparison to $Y_{\textsc{gg}}$, the roles of $\uc$, $\dc$ and of $\ec$, $\nc$ are interchanged.


\newpage

\section{Grand Unification for Dummies}
\label{app:guts_for_dummies}

In this section, we give some useful background information on Grand Unification. The branching rules and $\text{U}(1)$ charges that we present in Tab.~\ref{tab:U1charges_etc} may either be looked up in some standard reference on group theory \cite{Slansky:1981yr}, or calculated by hand using explicit expressions for the simple roots in some standard basis, see e.g.~Appendix 5A of Ref.~\cite{Green:1987sp}.

\begin{table}[h!]
\centering
\renewcommand{\arraystretch}{1.4}
\begin{tabular}{|c|c|c|c|c|c|c|c|c|}
\hline
\textsc{su(5) irrep}    & \textsc{sm irrep} & $U(1)_X$      & $U(1)_Y$      & $U(1)_{\textsc{B-L}}$   & $U(1)_R$   & $U(1)_{\text{fl}}$     & \textsc{flipped}      &   \textsc{gg}     \\
\hline
\hline
$\bs{10}$       & $(\bs{3}, \bs{2})$    & -$1$          & \phantom{-}$1/3$  & \phantom{-}$1/3$  & \phantom{-}$0$    & \phantom{-}$1/3$  & $Q$           &   $Q$         \\
            & $(\bsb{3}, \bs{1})$   & -$1$          & -$4/3$            & -$1/3$            & -$1$              & \phantom{-}$2/3$      & $\dc$         &   $\uc$       \\
            & $(\bs{1}, \bs{1})$    & -$1$          & \phantom{-}$2$    & \phantom{-}$1$    & \phantom{-}$1$    & \phantom{-}$0$    & $\nc$         &   $\ec$       \\
\hline
$\bsb{5}$       & $(\bsb{3}, \bs{1})$   & \phantom{-}$3$    & \phantom{-}$2/3$      & -$1/3$            &  \phantom{-}$1$       &          -$4/3$       & $\uc$         &   $\dc$       \\
            & $(\bs{1}, \bs{2})$    & \phantom{-}$3$    & -$1$              & -$1$              &  \phantom{-}$0$       & -$1$              & $L$           &   $L$         \\
\hline
$\bs{1}$        & $(\bs{1}, \bs{1})$    & -$5$          & \phantom{-}$0$    & \phantom{-}$1$    & -$1$          & \phantom{-}$2$    & $\ec$             &   $\nc$           \\
\hline
\end{tabular}
\caption{Branching rules and $\text{U}(1)$ charges for the matter multiplets in the Standard Model.}
\label{tab:U1charges_etc}
\end{table}


\subsection{Georgi-Glashow $\bs{\text{SU}(5)}$}
\label{sec:guts_georgi_glashow}

The matter content of the Standard Model fits in 3 irreducible representations,
\begin{equation}
\bs{10} \ra (\bs{3},\bs{2})_{1/3} + (\bsb{3},\bs{1})_{\text{-}4/3} + (\bs{1},\bs{1})_{2}, \qquad \bsb{5} \ra (\bsb{3},\bs{1})_{2/3} + (\bs{1},\bs{2})_{\text{-}1}, \qquad \bs{1} \ra (\bs{1},\bs{1})_{0},
\end{equation}
where we have indicated the charges under $\text{U}(1)_Y$ as subscripts. This hypercharge direction may be calculated by considering the Dynkin diagram associated with the symmetry breakdown:
\begin{center}
\begin{tabular}{ccccc}
{
\unitlength=.8mm
\begin{picture}(47,7)
\thicklines
\put(2, 5){\circle{4}} \put(17,5){\circle{4}} \put(32,5){\circle{4}}  \put(47,5){\circle{4}}
\put(4,5){\line(1,0){11}} \put(19,5){\line(1,0){11}} \put(34,5){\line(1,0){11}}
\put(0,-1){$\alpha_1$} \put(15,-1){$\alpha_2$} \put(30,-1){$\alpha_3$} \put(45,-1){$\alpha_4$}
\end{picture}
} &  \raisebox{2ex}[-3ex]{ $\bs{\longrightarrow}$  \hspace{-4ex}} & {
\unitlength=.8mm
\begin{picture}(47,7)
\thicklines
\put(2, 5){\circle{4}} \put(17,5){\circle{4}} \put(32,5){\circle{4}}  \put(47,5){\circle{4}}
\put(4,5){\line(1,0){11}} \put(19,5){\line(1,0){11}} \put(34,5){\line(1,0){11}}
\put(0,-1){$\alpha_1$} \put(15,-1){$\alpha_2$} \put(30,-1){$\alpha_3$} \put(45,-1){$\alpha_4$}
\put(29,2){\line(1,1){6}} \put(29,8){\line(1,-1){6}}
\end{picture}
}\\[2ex]
$\text{SU}(5)$  &  & $\text{SU}(3)_c \times \text{SU}(2)_L \times \text{U}(1)_Y$
\end{tabular}
\end{center}

$\text{U}(1)_Y$ is the dual of the root that is projected out. Keeping in mind that the dual roots are given in terms of the simple roots by the quadratic form matrix, which is the inverse of the Cartan matrix, we immediately obtain:
\begin{equation}
Y_{\textsc{gg}} = \frac{5}{3} \, \alpha_3^* = \frac{5}{3} \, \sum_{j=1}^4 (A_{\text{SU}(5)}^{-1})_{3j} \, \alpha_j = \frac{1}{3} \left( 2\alpha_1 + 4\alpha_2 + 6\alpha_3 + 3\alpha_4 \right)
\label{eq:Y_GG}
\end{equation}

The $\text{U}(1)_Y$ charges of the irreducible representations are calculated by taking the scalar product of the respective highest weight with $Y_{\textsc{gg}}$. We summarize the results in the 4th column of Tab.~\ref{tab:U1charges_etc}.


\newpage

\subsection{Pati-Salam}

The Standard Model particle content fits in one $\bs{16}$ of $\text{SO}(10)$. Under $\text{SO}(10) \ra \text{SU}(4)_c \times \text{SU}(2)_L \times \text{SU}(2)_R$, we have
\begin{equation}
\bs{16} \ra (\bs{4},\bs{2}, \bs{1}) + (\bsb{4},\bs{1}, \bs{2}).
\end{equation}
For more clarity, we break the Pati-Salam symmetry to that of the Standard Model in two steps. In the first step, we have
\begin{gather}
\text{SU}(4)_c \times \text{SU}(2)_L \times \text{SU}(2)_R \,\,\ra\,\, \text{SU}(3)_c \times \text{U}(1)_{\textsc{B-L}} \times \text{SU}(2)_L \times \text{SU}(2)_R\\[1ex]
(\bs{4},\bs{2}, \bs{1}) \ra  (\bs{3},\bs{2}, \bs{1})_{1/3} + (\bs{1},\bs{2}, \bs{1})_{\text{-}1}, \qquad (\bsb{4},\bs{1}, \bs{2}) \ra  (\bsb{3},\bs{1},\bs{2})_{\text{-}1/3} + (\bs{1},\bs{1},\bs{2})_{1},
\end{gather}
where the subscript denotes the $\text{U}(1)_{\textsc{B-L}}$ charge. In the second step, we have
\begin{gather}
\text{SU}(3)_c \times \text{U}(1)_{\textsc{B-L}} \times \text{SU}(2)_L \times \text{SU}(2)_R \ra \text{SU}(3)_c \times \text{U}(1)_{\textsc{B-L}} \times \text{SU}(2)_L \times \text{U}(1)_R\\[1ex]
(\bs{4},\bs{2}, \bs{1}) \ra  (\bs{3},\bs{2})_{1/3,0} + (\bs{1},\bs{2})_{\text{-}1,0}, \quad (\bsb{4},\bs{1}, \bs{2}) \ra  (\bsb{3},\bs{1})_{\text{-}1/3,1} + (\bsb{3},\bs{1})_{\text{-}1/3,\text{-}1} + (\bs{1},\bs{1})_{1,1} + (\bs{1},\bs{1})_{1,\text{-}1},\notag
\end{gather}
where the first subscript denotes the $\text{U}(1)_{\textsc{B-L}}$ charge as above and the second one is the $\text{U}(1)_{\text{R}}$ charge. Hypercharge is then given as the linear combination
\begin{equation}
\text{U}(1)_Y = \text{U}(1)_{\textsc{B-L}} + \text{U}(1)_R.
\label{eq:Y_PS}
\end{equation}

Again, this is most conveniently visualized in terms of Dynkin diagrams:

\begin{center}
\begin{tabular}{ccccc}
{
\unitlength=.8mm
\begin{picture}(47,21)
\thicklines
\put(2, 5){\circle{4}} \put(17,5){\circle{4}} \put(32,5){\circle{4}} \put(32,20){\circle{4}} \put(47,5){\circle{4}}
\put(4,5){\line(1,0){11}} \put(19,5){\line(1,0){11}} \put(32,7){\line(0,1){11}} \put(34,5){\line(1,0){11}}
\put(0,-1){$\alpha_1$} \put(15,-1){$\alpha_2$} \put(30,-1){$\alpha_3$} \put(35,19){$\alpha_5$} \put(45,-1){$\alpha_4$}
\end{picture}
} &  \raisebox{2ex}[-3ex]{ $\bs{\longrightarrow}$ \hspace{-4ex}} & {
\unitlength=.8mm
\begin{picture}(45,21)
\thicklines
\put(2, 5){\circle{4}} \put(17,5){\circle{4}} \put(17,20){\circle{4}} \put(32,5){\circle{4}} \put(32,20){\circle{4}} \put(47,5){\circle{4}}
\put(4,5){\line(1,0){11}} \put(17,7){\dashline{2}(0,0)(0,11)} \put(19,5){\line(1,0){11}} \put(32,7){\line(0,1){11}} \put(34,5){\line(1,0){11}}
\put(0,-1){$\alpha_1$} \put(15,-1){$\alpha_2$} \put(20,19){$\alpha_0$} \put(30,-1){$\alpha_3$} \put(35,19){$\alpha_5$} \put(45,-1){$\alpha_4$}
\put(29,2){\line(1,1){6}} \put(29,8){\line(1,-1){6}}
\end{picture}
} &  \raisebox{2ex}[-3ex]{ $\bs{\longrightarrow}$  \hspace{-4ex}} & {
\unitlength=.8mm
\begin{picture}(62,21)
\thicklines
\put(2, 5){\circle{4}} \put(17,5){\circle{4}} \put(32,5){\circle{4}} \put(47,5){\circle{4}} \put(62,5){\circle{4}}
\put(4,5){\line(1,0){11}} \put(19,5){\dashline{2}(0,0)(11,0)}
\put(0,-1){$\alpha_1$} \put(15,-1){$\alpha_2$} \put(30,-1){$\alpha_0$} \put(45,-1){$\alpha_5$} \put(60,-1){$\alpha_4$}
\put(29,2){\line(1,1){6}} \put(29,8){\line(1,-1){6}}
\put(59,2){\line(1,1){6}} \put(59,8){\line(1,-1){6}}
\end{picture}
}\\[3ex]
$\text{SO}(10)$ &  & $\text{SU}(4)_c \times \text{SU}(2)_L \times \text{SU}(2)_R$ &  & $\text{SU}(3)_c \times \text{SU}(2)_L \times \text{U}(1)_{\textsc{B-L}} \times \text{U}(1)_R$
\end{tabular}
\end{center}
We can easily calculate the directions corresponding to the two abelian factors:
\begin{gather}
Y_{\textsc{b-l}} = \frac{4}{3} \, \alpha_0^* = \frac{4}{3} \, \sum_{j=1,2,0} (A_{\text{SU}(4)}^{-1})_{0j} \, \alpha_j = \frac{1}{3} \left( \alpha_1 + 2\alpha_2 + 3\alpha_0 \right)\\[1ex]
Y_{\textsc{r}} = \alpha_4^* = \sum_{j=4} (A_{\text{SU}(2)}^{-1})_{4j} \, \alpha_j = \frac{1}{2} \, \alpha_4
\end{gather}

Incidentally, we should remark that
\begin{equation}
Y_{\textsc{gg}} = Y_{\textsc{b-l}} + Y_{\textsc{r}},
\end{equation}
as was to be expected.


\newpage

\subsection{Flipped $\bs{\text{SU}(5)}$}

Under $\text{SO}(10) \ra \text{SU}(5) \times \text{U}(1)_X$ we have
\begin{equation}
\bs{16} \ra \bs{10}_{\text{-}1} + \bsb{5}_{3} + \bs{1}_{\text{-}5}.
\end{equation}
(This also defines what we mean by $\text{U}(1)_X$.) Breaking the $\text{SU}(5)$ symmetry to that of the Standard Model, we obtain
\begin{equation}
\bs{10} \ra (\bs{3},\bs{2})_{\text{-}1, 1/3} + (\bsb{3},\bs{1})_{\text{-}1, \text{-}4/3} + (\bs{1},\bs{1})_{\text{-}1, 2}, \quad \bsb{5} \ra (\bsb{3},\bs{1})_{3,2/3} + (\bs{1},\bs{2})_{3,\text{-}1}, \quad \bs{1} \ra (\bs{1},\bs{1})_{\text{-}5, 0},
\end{equation}
where the the first and second subscript denotes the $\text{U}(1)_X$ and $\text{U}(1)_Y$ charge, respectively. In terms of Dynkin diagrams, this reads:
\begin{center}
\begin{tabular}{ccccc}
{
\unitlength=.8mm
\begin{picture}(47,21)
\thicklines
\put(2, 5){\circle{4}} \put(17,5){\circle{4}} \put(32,5){\circle{4}} \put(32,20){\circle{4}} \put(47,5){\circle{4}}
\put(4,5){\line(1,0){11}} \put(19,5){\line(1,0){11}} \put(32,7){\line(0,1){11}} \put(34,5){\line(1,0){11}}
\put(0,-1){$\alpha_1$} \put(15,-1){$\alpha_2$} \put(30,-1){$\alpha_3$} \put(35,19){$\alpha_5$} \put(45,-1){$\alpha_4$}
\end{picture}
} &  \raisebox{4ex}[-3ex]{ $\bs{\longrightarrow}$ } & {
\unitlength=.8mm
\begin{picture}(47,21)
\thicklines
\put(2, 5){\circle{4}} \put(17,5){\circle{4}} \put(32,5){\circle{4}} \put(32,20){\circle{4}} \put(47,5){\circle{4}}
\put(4,5){\line(1,0){11}} \put(19,5){\line(1,0){11}} \put(32,7){\line(0,1){11}} \put(34,5){\line(1,0){11}}
\put(0,-1){$\alpha_1$} \put(15,-1){$\alpha_2$} \put(30,-1){$\alpha_3$} \put(35,19){$\alpha_5$} \put(45,-1){$\alpha_4$}
\put(29,17){\line(1,1){6}} \put(29,23){\line(1,-1){6}}
\end{picture}
} &  \raisebox{4ex}[-3ex]{ $\bs{\longrightarrow}$  \hspace{-4ex}} & {
\unitlength=.8mm
\begin{picture}(47,21)
\thicklines
\put(2, 5){\circle{4}} \put(17,5){\circle{4}} \put(32,5){\circle{4}} \put(32,20){\circle{4}} \put(47,5){\circle{4}}
\put(4,5){\line(1,0){11}} \put(19,5){\line(1,0){11}} \put(32,7){\line(0,1){11}} \put(34,5){\line(1,0){11}}
\put(0,-1){$\alpha_1$} \put(15,-1){$\alpha_2$} \put(30,-1){$\alpha_3$} \put(35,19){$\alpha_5$} \put(45,-1){$\alpha_4$}
\put(29,2){\line(1,1){6}} \put(29,8){\line(1,-1){6}}
\put(29,17){\line(1,1){6}} \put(29,23){\line(1,-1){6}}
\end{picture}
}\\[3ex]
$\text{SO}(10)$ &  & $\text{SU}(5) \times \text{U}(1)_X$  &  & $\text{SU}(3)_c \times \text{SU}(2)_L \times \text{U}(1)_X \times \text{U}(1)_Y$
\end{tabular}
\end{center}

The direction corresponding to $\text{U}(1)_X$ is given by:
\begin{equation}
\text{Y}_X = 4 \, \alpha_5^* = 4 \, \sum_{j=1}^5 (A_{\text{SO}(10)}^{-1})_{5j} \, \alpha_j = \left( 2\alpha_1 + 4\alpha_2 + 6\alpha_3 + 3\alpha_4  + 5\alpha_5 \right)
\end{equation}

In flipped $\text{SU}(5)$, hypercharge is a linear combination of this direction and $\text{B-L}$, namely:
\begin{equation}
\text{Y}_{\text{fl}} = -\frac{1}{2} \text{Y}_X - \frac{1}{2} \text{Y}_\textsc{B-L}
\end{equation}
Calculating the hypercharge assignments for the elementary particles, it turns out that in comparison to Georgi-Glashow $\text{SU}(5)$,  the roles of $\uc$, $\dc$ and of $\ec$, $\nc$ are interchanged, cf.~Tab.~\ref{tab:U1charges_etc}.


\newpage

\section{Georgi-Glashow Hypercharge in Disguise}
\label{sec:hypercharge_outside_SO10}

Consider the following two models below, to which we will refer as

\bigskip

{\bfseries First Model}\\[-4ex]
\begin{center}
\renewcommand{\arraystretch}{1.4}
\begin{tabular}{cccccccccccccccccc}
$V_6$     & = &  \big(  $\tfrac{1}{3}$ &     $\tfrac{1}{2}$ &    \text{-}$\tfrac{1}{3}$ &    \text{-}$\tfrac{1}{6}$ &    \text{-}$\tfrac{1}{6}$ &    \text{-}$\tfrac{1}{6}$ &    \text{-}$\tfrac{1}{6}$ &    \text{-}$\tfrac{1}{6}$ &     $\tfrac{1}{12}$ &     $\tfrac{1}{4}$ &     $\tfrac{1}{12}$ &    \text{-}$\tfrac{1}{12}$ &    \text{-}$\tfrac{1}{12}$ &    \text{-}$\tfrac{1}{12}$ &    $\tfrac{11}{12}$ &    $\tfrac{11}{12}$  \big)\\
$W_3$     & = & \big(      \text{-}$\tfrac{1}{6}$ &     $\tfrac{1}{2}$ &     $\tfrac{1}{6}$ &    \text{-}$\tfrac{1}{6}$ &    \text{-}$\tfrac{1}{6}$ &     $\tfrac{1}{6}$ &     $\tfrac{7}{6}$ &     $\tfrac{7}{6}$ &     $\tfrac{7}{6}$ &    \text{-}$\tfrac{5}{6}$ &    \text{-}$\tfrac{5}{6}$ &    \text{-}$\tfrac{5}{6}$ &    \text{-}$\tfrac{5}{6}$ &    \text{-}$\tfrac{5}{6}$ &  \text{-}$\tfrac{25}{6}$ &  \text{-}$\tfrac{13}{6}$  \big)\\
$W_2$     & = &  \big(      \text{-}$\tfrac{3}{4}$ &    \text{-}$\tfrac{1}{4}$ &    \text{-}$\tfrac{1}{4}$ &    \text{-}$\tfrac{1}{4}$ &    \text{-}$\tfrac{1}{4}$ &     $\tfrac{1}{4}$ &     $\tfrac{5}{4}$ &     $\tfrac{5}{4}$ &     $\tfrac{3}{4}$ &     $\tfrac{1}{4}$ &    \text{-}$\tfrac{1}{4}$ &    \text{-}$\tfrac{3}{4}$ &    \text{-}$\tfrac{3}{4}$ &    \text{-}$\tfrac{3}{4}$ &    $\tfrac{25}{4}$ &    $\tfrac{29}{4}$  \big)\\
\end{tabular}
\end{center}

\medskip

{\bfseries Second Model}\\[-4ex]
\begin{center}
\renewcommand{\arraystretch}{1.4}
\begin{tabular}{cccccccccccccccccc}
$V_6$     & = &  \big(      $\tfrac{2}{3}$ &     $\tfrac{1}{3}$ &     $\tfrac{1}{6}$ &     $\tfrac{1}{6}$ &     0   &   0  &    0   &   0  &   \text{-}$\tfrac{1}{6}$ &     $\tfrac{1}{2}$ &     $\tfrac{1}{2}$ &     $\tfrac{1}{2}$ &     $\tfrac{1}{2}$ &     $\tfrac{1}{2}$ &    \text{-}$\tfrac{1}{2}$ &     $\tfrac{1}{2}$ \big)\\
$W_3$     & = &  \big(      $\tfrac{1}{6}$ &    \text{-}$\tfrac{1}{6}$ &     $\tfrac{1}{6}$ &     $\tfrac{1}{6}$ &    \text{-}$\tfrac{1}{2}$ &     $\tfrac{1}{6}$ &     $\tfrac{1}{6}$ &    \text{-}$\tfrac{1}{6}$ &    \text{-}$\tfrac{1}{3}$ &     $\tfrac{2}{3}$ &     $\tfrac{2}{3}$ &     $\tfrac{1}{3}$ &     $\tfrac{1}{3}$ &     $\tfrac{1}{3}$ &     $\tfrac{1}{3}$ &    \text{-}$\tfrac{1}{3}$ \big)\\
$W_2$     & = &  \big(     \text{-}$\tfrac{3}{4}$ &     $\tfrac{1}{4}$ &     $\tfrac{3}{4}$ &    \text{-}$\tfrac{3}{4}$ &    \text{-}$\tfrac{1}{4}$ &    \text{-}$\tfrac{1}{4}$ &    \text{-}$\tfrac{1}{4}$ &     $\tfrac{1}{4}$ &    \text{-}$\tfrac{1}{4}$ &    \text{-}$\tfrac{1}{4}$ &    \text{-}$\tfrac{1}{4}$ &    \text{-}$\tfrac{3}{4}$ &    \text{-}$\tfrac{3}{4}$ &    \text{-}$\tfrac{3}{4}$ &     $\tfrac{1}{4}$ &    \text{-}$\tfrac{1}{4}$ \big)\\
\end{tabular}
\end{center}

\medskip

for the sake of brevity. The gauge group in {\it both cases} is
\begin{equation}
\text{SU}(3) \times \text{SU}(2) \times \text{SU}(2) \times \text{U}(1)^4 \times \text{SU}(7)' \times \text{U}(1)'^2,
\end{equation}

and their {\it spectra coincide:}

\begin{center}
\normalsize
\renewcommand{\arraystretch}{1.4}
\begin{tabular}{|r|r|r|r|r|}
\hline
$3\times(\bs{3},\bs{2},\bs{1},\bs{1})$ & $3\times(\bsb{3},\bs{1},\bs{2},\bs{1})$    & $2\times(\bs{1},\bs{2},\bs{2},\bs{1})$  & $4\times(\bs{1},\bs{1},\bs{1},\bs{7})$  &   $34\times(\bs{1},\bs{1},\bs{1},\bs{1})$\\
\hline
                                        & $12\times(\bs{3},\bs{1},\bs{1},\bs{1})$   & $13\times(\bs{1},\bs{2},\bs{1},\bs{1})$ & $4\times(\bs{1},\bs{1},\bs{1},\bsb{7})$ &   \\
\cline{2-4}
                                        & $12\times(\bsb{3},\bs{1},\bs{1},\bs{1})$  & $17\times(\bs{1},\bs{1},\bs{2},\bs{1})$ &                                         &   \\
\hline
\end{tabular}
\end{center}

\medskip

In both cases, the shift $V_6$ breaks
\begin{equation}
\text{E}_8 \ra \text{SO}(10) \times \text{SU}(2) \times \text{SU}(2).
\end{equation}

For the first model, the Wilson lines break to:

\bigskip
\bigskip

\begin{tabular}{ccc}
{
\unitlength=.8mm
\begin{picture}(75,21)
\thicklines
\put(2, 5){\circle{4}} \put(17,5){\circle{4}} \put(17,20){\circle{4}} \put(32,5){\circle{4}} \put(32,20){\circle{4}} \put(47,5){\circle{4}} \put(62,5){\circle{4}} \put(77, 5){\circle{4}}
\put(4,5){\line(1,0){11}} \put(17,7){\dashline{2}(0,0)(0,11)} \put(19,5){\line(1,0){11}} \put(32,7){\line(0,1){11}} \put(34,5){\line(1,0){11}}
\put(0,-1){$\alpha_1$} \put(15,-1){$\alpha_2$} \put(20,19){$\alpha_0$} \put(30,-1){$\alpha_3$} \put(35,19){$\alpha_5$} \put(45,-1){$\alpha_4$} \put(60,-1){$\alpha_6$} \put(75,-1){$\alpha_7$}
\end{picture}
} &  \raisebox{4ex}[-3ex]{\qquad $\bs{\longrightarrow}$ \quad} & {
\unitlength=.8mm
\begin{picture}(75,21)
\thicklines
\put(2, 5){\circle{4}} \put(17,5){\circle{4}} \put(17,20){\circle{4}} \put(32,5){\circle{4}} \put(32,20){\circle{4}} \put(47,5){\circle{4}} \put(62,5){\circle{4}} \put(77, 5){\circle{4}}
\put(4,5){\line(1,0){11}} \put(17,7){\dashline{2}(0,0)(0,11)} \put(19,5){\line(1,0){11}} \put(32,7){\line(0,1){11}} \put(34,5){\line(1,0){11}}
\put(0,-1){$\alpha_1$} \put(15,-1){$\alpha_2$} \put(20,19){$\alpha_0$} \put(30,-1){$\alpha_3$} \put(35,19){$\alpha_5$} \put(45,-1){$\alpha_4$} \put(60,-1){$\alpha_6$} \put(75,-1){$\alpha_7$}
\put(29,2){\line(1,1){6}} \put(29,8){\line(1,-1){6}}
\put(29,17){\line(1,1){6}} \put(29,23){\line(1,-1){6}}
\put(14,17){\line(1,1){6}} \put(14,23){\line(1,-1){6}}
\put(59,2){\line(1,1){6}} \put(59,8){\line(1,-1){6}}
\end{picture}
}\\[3ex]
$\uwave{\text{SO}(10)} \times \text{SU}(2) \times \text{SU}(2)$ &  & $\uwave{\text{SU}(3) \times \text{SU}(2) \times \text{U}(1)^2} \times \text{SU}(2) \times \text{U}(1)$
\end{tabular}

\bigskip

The underlining indicates the breaking of $\text{SO}(10)$.

\newpage

In contrast, in the second model, the Wilson lines break to:\\[2ex]

\begin{tabular}{ccc}
{
\unitlength=.8mm
\begin{picture}(75,21)
\thicklines
\put(2, 5){\circle{4}} \put(17,5){\circle{4}} \put(17,20){\circle{4}} \put(32,5){\circle{4}} \put(32,20){\circle{4}} \put(47,5){\circle{4}} \put(62,5){\circle{4}} \put(77, 5){\circle{4}}
\put(4,5){\line(1,0){11}} \put(17,7){\dashline{2}(0,0)(0,11)} \put(19,5){\line(1,0){11}} \put(32,7){\line(0,1){11}} \put(34,5){\line(1,0){11}}
\put(0,-1){$\alpha_6$} \put(15,-1){$\alpha_1$} \put(20,19){$\alpha_0$} \put(30,-1){$\alpha_3$} \put(35,19){$\alpha_2$} \put(45,-1){$\alpha_4$} \put(60,-1){$\alpha_5$} \put(75,-1){$\alpha_7$}
\end{picture}
} &  \raisebox{4ex}[-3ex]{\qquad $\bs{\longrightarrow}$ \quad} & {
\unitlength=.8mm
\begin{picture}(75,21)
\thicklines
\put(2, 5){\circle{4}} \put(17,5){\circle{4}} \put(17,20){\circle{4}} \put(32,5){\circle{4}} \put(32,20){\circle{4}} \put(47,5){\circle{4}} \put(62,5){\circle{4}} \put(77, 5){\circle{4}}
\put(4,5){\line(1,0){11}} \put(17,7){\dashline{2}(0,0)(0,11)} \put(19,5){\line(1,0){11}} \put(32,7){\line(0,1){11}} \put(34,5){\line(1,0){11}}
\put(0,-1){$\alpha_6$} \put(15,-1){$\alpha_1$} \put(20,19){$\alpha_0$} \put(30,-1){$\alpha_3$} \put(35,19){$\alpha_2$} \put(45,-1){$\alpha_4$} \put(60,-1){$\alpha_5$} \put(75,-1){$\alpha_7$}
\put(14,2){\line(1,1){6}} \put(14,8){\line(1,-1){6}}
\put(44,2){\line(1,1){6}} \put(44,8){\line(1,-1){6}}
\put(59,2){\line(1,1){6}} \put(59,8){\line(1,-1){6}}
\put(74,2){\line(1,1){6}} \put(74,8){\line(1,-1){6}}
\end{picture}
}\\[3ex]
$\uwave{\text{SO}(10)} \times \text{SU}(2) \times \text{SU}(2)$ &  & $\uwave{\text{SU}(3) \times \text{SU}(2) \times \text{SU}(2) \times \text{U}(1)} \times \text{U}(1)^2$
\end{tabular}

\bigskip

In both cases, the resulting gauge group and the spectra are the same.
The difference lies in the way how the
$\text{SO}(10)$ breaks. For the first model, hypercharge that lies in
$\text{SO}(10)$ (see Appendix \ref{app:guts_for_dummies}) gives a
vector-like spectrum, whereas for the second model, we cannot even attempt
to construct this hypercharge direction, since the intermediate
$\text{SO}(10)$ is seemingly missing. Our general ansatz will yield a hypercharge
direction in either case, but for the second model, the connection between the underlying
$\text{SO}(10)$ and hypercharge is disguised. This complication is {\it not relevant} for the mini-landscape
search, since there, we {\it first} constructed the hypercharge direction and {\it then} identified the models with
identical non-abelian spectra.


\newpage

\section{Georgi-Glashow Unification without $\bs{\text{SU}(5)}$}
\label{app:details_weird_model}

Consider the following model with shift and Wilson lines

\begin{center}
\renewcommand{\arraystretch}{1.4}
\begin{tabular}{cccccccccccccccccc}
 $V_6$     & = &  \big(    $\frac{2}{3}$ &    $\frac{1}{3}$ &    $\frac{1}{6}$ &    $\frac{1}{6}$ &    $0$ &    $0$ &    $0$ &    $0$ &    $\frac{1}{3}$ &    $0$ &    $0$ &    $0$ &    $0$ &    $0$ &    $0$ &    $0$  \big)\\
 $W_3$     & = &  \big(    $\frac{1}{6}$ &    $\frac{1}{6}$ &   -$\frac{1}{6}$ &   -$\frac{1}{2}$ &    $\frac{1}{2}$ &    $\frac{1}{6}$ &    $\frac{1}{6}$ &    $\frac{1}{6}$ &   -$\frac{2}{3}$ &    $\frac{1}{3}$ &    $\frac{1}{3}$ &    $0$ &    $0$ &    $0$ &    $0$ &    $0$  \big)\\
$W_2$     & = &  \big(    -$\frac{1}{4}$ &    $\frac{1}{4}$ &    $\frac{1}{4}$ &    $\frac{1}{4}$ &    $\frac{1}{4}$ &   -$\frac{1}{4}$ &   -$\frac{1}{4}$ &   -$\frac{1}{4}$ &    $0$ &   -$\frac{1}{2}$ &    $\frac{1}{2}$ &    $0$ &    $0$ &    $0$ &    $0$ &    $0$  \big)\\
\end{tabular}
\end{center}
and hypercharge
\begin{center}
\renewcommand{\arraystretch}{1.4}
\begin{tabular}{cccccccccccccccccc}
$Y$     & = &  \big(    $\frac{1}{2}$ &   -$\frac{1}{2}$ &   -$\frac{1}{2}$ &   -$\frac{1}{2}$ &    $\frac{3}{2}$ &   -$\frac{1}{6}$ &   -$\frac{1}{6}$ &   -$\frac{1}{6}$ &    $0$ &    $0$ &    $0$ &    $0$ &    $0$ &    $0$ &    $0$ &    $0$  \big).\\
\end{tabular}
\end{center}

The shift and Wilson lines break
\begin{equation*}
\begin{split}
\text{E}_8 \times \text{E}_8' \quad \overset{V}{\longrightarrow} \quad \text{SO}&(10) \times \text{SU}(2) \times \text{SU}(2) \times \text{SO}(14)'\\
\quad & \overset{W_3}{\longrightarrow} \quad \text{SU}(4) \times \text{SO}(10)' \times \text{SU}(2)' \quad \overset{W_2}{\longrightarrow} \quad \text{SU}(3) \times \text{SO}(10)' \times \text{SU}(2)'.
\end{split}
\end{equation*}
The gauge groups for the color and weak interactions are located in the first and second $\text{E}_8$, respectively, whereas hypercharge lies fully in the first one. As such, the matter representations like $(\bs{3}, \bs{2})$ cannot originate from the $\bs{16}$ of $\text{SO}(10)$, since $\text{SU}(3)\times\text{SU}(2) \not\subset \text{SO}(10)$ in the first place.

\begin{figure}[h!]
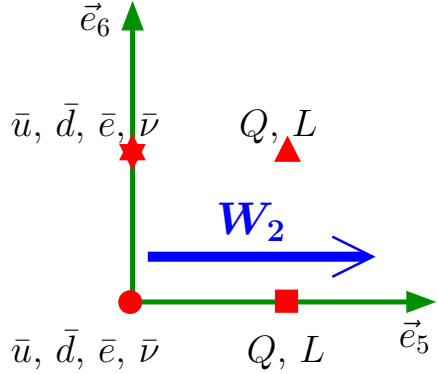

\centering
\begin{center}\input ./graphs/z6-II-T1.pstex_t \end{center}
\caption{Localization of the matter fields in the 3rd torus.}
\label{fig:localization_matter}
\end{figure}

\medskip

Nevertheless, evaluating $Y$ on the $(\bs{16}, \bs{1}, \bs{1}, \bs{1})$ which lives at the origin (for $W_3 = W_2 = 0$), we recover the standard charges. A quick calculation shows that $Y$ lies in $\text{SO}(10)$, so it corresponds to either regular or flipped $\text{SU}(5)$ for some choice of simple roots of $\text{SO}(10)$.

\medskip

It is still true that the $\bs{16}$ is unaffected by the orbifold projections, so the massless states which we expect to survive are
\begin{equation*}
\bs{16} \rightarrow (\bs{3},\bs{1})_{1/3} + (\bs{3},\bs{1})_{1/3} + (\bsb{3},\bs{1})_{\text{-}4/3} + (\bsb{3},\bs{1})_{2/3} + (\bs{1},\bs{1})_{\text{-}1} + (\bs{1},\bs{1})_{\text{-}1} + (\bs{1},\bs{1})_{2} + (\bs{1},\bs{1})_{0}
\end{equation*}
at the origin in the 5th twisted sector, and a quick check shows that this is indeed true. The big difference is that only $\uc$, $\dc$, $\ec$ and $\nc$   (see Tab.~\ref{tab:spectrum_SM} on page \pageref{tab:spectrum_SM}) live in this $\bs{16}$ at the origin, and the other representations in the multiplet have an interpretation as exotic particles. Since there is no Wilson line along the $\vec{e}_6$ in the torus, the spectrum is degenerate along this direction.

\medskip

In Fig.~\ref{fig:localization_matter}, we present the localization of the Standard Model fields. There are $2 \times (\bs{3},\bs{2})_{1/3}$ in the 5th twisted sector living at ${\textstyle \color{red}{\blacktriangle}}$ and ${\scriptstyle \color{red}{\blacksquare}}$\,, respectively. Their highest weights are
\begin{center}
\renewcommand{\arraystretch}{1.4}
\begin{tabular}{lccccccccccccccccc}
$Q_{\textstyle \blacktriangle}$     & = &  \big(    $\frac{1}{12}$ &   -$\frac{1}{12}$ &    $\frac{1}{12}$ &    $\frac{1}{12}$ &    $\frac{1}{4}$ &    $\frac{3}{4}$ &   -$\frac{1}{4}$ &    -$\frac{1}{4}$ &    -$\frac{1}{3}$ &     $\frac{1}{2}$ &    -$\frac{1}{2}$ &       $0$ &       $0$ &       $0$ &       $0$ &       $0$  \big),\\
$Q_{\scriptstyle \blacksquare}$     & = &  \big(    $\frac{1}{12}$ &   -$\frac{1}{12}$ &    $\frac{1}{12}$ &    $\frac{1}{12}$ &    $\frac{1}{4}$ &    $\frac{3}{4}$ &   -$\frac{1}{4}$ &    -$\frac{1}{4}$ &    -$\frac{1}{3}$ &     $\frac{1}{2}$ &    -$\frac{1}{2}$ &       $0$ &       $0$ &       $0$ &       $0$ &       $0$  \big).\\
\end{tabular}
\end{center}
These states are not part of a $\bs{16}$ or any other $\text{SO}(10)$ irreducible representation, as can be seen by calculating the Dynkin labels w.r.t.~the simple roots. Nevertheless, the scalar product with $Y$ gives the correct hypercharge for a left-handed quark doublet in both cases.


\clearpage
\newpage

\bibliography{mybibliography}

\bibliographystyle{./utphys}

\end{document}

%% file: graphs/firsttwisted.pstex_t
\begin{picture}(0,0)%
\includegraphics{firsttwisted.pstex}%
\end{picture}%
\setlength{\unitlength}{2486sp}%
\begingroup\makeatletter\ifx\SetFigFont\undefined%
\gdef\SetFigFont#1#2#3#4#5{%
  \reset@font\fontsize{#1}{#2pt}%
  \fontfamily{#3}\fontseries{#4}\fontshape{#5}%
  \selectfont}%
\fi\endgroup%
\begin{picture}(9558,2271)(826,-2749)
\put(2071,-1366){\makebox(0,0)[lb]{\smash{{\SetFigFont{9}{10.8}{\rmdefault}{\mddefault}{\updefault}{\color[rgb]{0,0,0}$\vec{e}_2$}%
}}}}
\put(1711,-2581){\makebox(0,0)[lb]{\smash{{\SetFigFont{9}{10.8}{\rmdefault}{\mddefault}{\updefault}{\color[rgb]{0,0,0}$\vec{e}_1$}%
}}}}
\put(6301,-2581){\makebox(0,0)[lb]{\smash{{\SetFigFont{9}{10.8}{\rmdefault}{\mddefault}{\updefault}{\color[rgb]{0,0,0}$\vec{e}_3$}%
}}}}
\put(10126,-2581){\makebox(0,0)[lb]{\smash{{\SetFigFont{9}{10.8}{\rmdefault}{\mddefault}{\updefault}{\color[rgb]{0,0,0}$\vec{e}_5$}%
}}}}
\put(8236,-691){\makebox(0,0)[lb]{\smash{{\SetFigFont{9}{10.8}{\rmdefault}{\mddefault}{\updefault}{\color[rgb]{0,0,0}$\vec{e}_6$}%
}}}}
\put(8326,-2641){\makebox(0,0)[lb]{\smash{{\SetFigFont{14}{16.8}{\rmdefault}{\mddefault}{\updefault}{\color[rgb]{0,0,0}$\bs{16}$}%
}}}}
\put(9046,-1981){\makebox(0,0)[lb]{\smash{{\SetFigFont{11}{13.2}{\rmdefault}{\mddefault}{\updefault}{\color[rgb]{0,0,1}$\bs{W_2}$}%
}}}}
\put(3571,-946){\makebox(0,0)[lb]{\smash{{\SetFigFont{9}{10.8}{\rmdefault}{\mddefault}{\updefault}{\color[rgb]{0,0,0}$\vec{e}_4$}%
}}}}
\put(4876,-1951){\makebox(0,0)[lb]{\smash{{\SetFigFont{11}{13.2}{\rmdefault}{\mddefault}{\updefault}{\color[rgb]{0,0,1}$\bs{W_3}$}%
}}}}
\put(8341,-1231){\makebox(0,0)[lb]{\smash{{\SetFigFont{14}{16.8}{\rmdefault}{\mddefault}{\updefault}{\color[rgb]{0,0,0}$\bs{16}$}%
}}}}
\put(8416,-1501){\makebox(0,0)[lb]{\smash{{\SetFigFont{20}{24.0}{\rmdefault}{\mddefault}{\updefault}{\color[rgb]{0,0,0}$\color{red}{\star}$}%
}}}}
\end{picture}%

%% file: graphs/z6-II-T1.pstex_t
\begin{picture}(0,0)%
\includegraphics{z6-II-T1.pstex}%
\end{picture}%
\setlength{\unitlength}{4144sp}%
\begingroup\makeatletter\ifx\SetFigFont\undefined%
\gdef\SetFigFont#1#2#3#4#5{%
  \reset@font\fontsize{#1}{#2pt}%
  \fontfamily{#3}\fontseries{#4}\fontshape{#5}%
  \selectfont}%
\fi\endgroup%
\begin{picture}(2553,2260)(7831,-2738)
\put(10126,-2581){\makebox(0,0)[lb]{\smash{{\SetFigFont{14}{16.8}{\rmdefault}{\mddefault}{\updefault}{\color[rgb]{0,0,0}$\vec{e}_5$}%
}}}}
\put(8236,-691){\makebox(0,0)[lb]{\smash{{\SetFigFont{14}{16.8}{\rmdefault}{\mddefault}{\updefault}{\color[rgb]{0,0,0}$\vec{e}_6$}%
}}}}
\put(9226,-2671){\makebox(0,0)[lb]{\smash{{\SetFigFont{14}{16.8}{\rmdefault}{\mddefault}{\updefault}{\color[rgb]{0,0,0}$Q$, $L$}%
}}}}
\put(7831,-1321){\makebox(0,0)[lb]{\smash{{\SetFigFont{14}{16.8}{\rmdefault}{\mddefault}{\updefault}{\color[rgb]{0,0,0}$\uc$, $\dc$, $\ec$, $\nc$}%
}}}}
\put(7831,-2671){\makebox(0,0)[lb]{\smash{{\SetFigFont{14}{16.8}{\rmdefault}{\mddefault}{\updefault}{\color[rgb]{0,0,0}$\uc$, $\dc$, $\ec$, $\nc$}%
}}}}
\put(9181,-1321){\makebox(0,0)[lb]{\smash{{\SetFigFont{14}{16.8}{\rmdefault}{\mddefault}{\updefault}{\color[rgb]{0,0,0}$Q$, $L$}%
}}}}
\put(9046,-1906){\makebox(0,0)[lb]{\smash{{\SetFigFont{17}{20.4}{\rmdefault}{\mddefault}{\updefault}{\color[rgb]{0,0,1}$\bs{W_2}$}%
}}}}
\end{picture}%

%% file: paper.bbl
\providecommand{\href}[2]{#2}\begingroup\raggedright\begin{thebibliography}{10}

\bibitem{Lebedev:2006kn}
O.~Lebedev {\em et al.}, ``A mini-landscape of exact mssm spectra in heterotic
  orbifolds,'' {\em Phys. Lett.} {\bf B645} (2007) 88--94,
\href{http://www.arXiv.org/abs/hep-th/0611095}{{\tt hep-th/0611095}}.

\bibitem{Kobayashi:2004ud}
T.~Kobayashi, S.~Raby, and R.-J. Zhang, ``Constructing 5d orbifold grand
  unified theories from heterotic strings,'' {\em Phys. Lett.} {\bf B593}
  (2004) 262--270,
\href{http://www.arXiv.org/abs/hep-ph/0403065}{{\tt hep-ph/0403065}}.

\bibitem{Forste:2004ie}
S.~F{\"{o}}rste, H.~P. Nilles, P.~K.~S. Vaudrevange, and A.~Wingerter,
  ``Heterotic brane world,'' {\em Phys. Rev.} {\bf D70} (2004) 106008,
\href{http://www.arXiv.org/abs/hep-th/0406208}{{\tt hep-th/0406208}}.

\bibitem{Kobayashi:2004ya}
T.~Kobayashi, S.~Raby, and R.-J. Zhang, ``Searching for realistic 4d string
  models with a pati-salam symmetry: Orbifold grand unified theories from
  heterotic string compactification on a {${\mathbb Z}_6$} orbifold,'' {\em
  Nucl. Phys.} {\bf B704} (2005) 3--55,
\href{http://www.arXiv.org/abs/hep-ph/0409098}{{\tt hep-ph/0409098}}.

\bibitem{Buchmuller:2006ik}
W.~Buchm{\"{u}}ller, K.~Hamaguchi, O.~Lebedev, and M.~Ratz, ``Supersymmetric
  standard model from the heterotic string {II},''
\href{http://www.arXiv.org/abs/hep-th/0606187}{{\tt hep-th/0606187}}.

\bibitem{Buchmuller:2004hv}
W.~Buchm{\"{u}}ller, K.~Hamaguchi, O.~Lebedev, and M.~Ratz, ``Dual models of
  gauge unification in various dimensions,'' {\em Nucl. Phys.} {\bf B712}
  (2005) 139--156,
\href{http://www.arXiv.org/abs/hep-ph/0412318}{{\tt hep-ph/0412318}}.

\bibitem{Buchmuller:2005jr}
W.~Buchm{\"{u}}ller, K.~Hamaguchi, O.~Lebedev, and M.~Ratz, ``Supersymmetric
  standard model from the heterotic string,'' {\em Phys. Rev. Lett.} {\bf 96}
  (2006) 121602,
\href{http://www.arXiv.org/abs/hep-ph/0511035}{{\tt hep-ph/0511035}}.

\bibitem{Buchmuller:2005sh}
W.~Buchm{\"{u}}ller, K.~Hamaguchi, O.~Lebedev, and M.~Ratz, ``Local grand
  unification,''
\href{http://www.arXiv.org/abs/hep-ph/0512326}{{\tt hep-ph/0512326}}.

\bibitem{Forste:2005rs}
S.~F{\"{o}}rste, H.~P. Nilles, and A.~Wingerter, ``Geometry of rank
  reduction,'' {\em Phys. Rev.} {\bf D72} (2005) 026001,
\href{http://www.arXiv.org/abs/hep-th/0504117}{{\tt hep-th/0504117}}.

\bibitem{Forste:2005gc}
S.~F{\"{o}}rste, H.~P. Nilles, and A.~Wingerter, ``The higgs mechanism in
  heterotic orbifolds,'' {\em Phys. Rev.} {\bf D73} (2006) 066011,
\href{http://www.arXiv.org/abs/hep-th/0512270}{{\tt hep-th/0512270}}.

\bibitem{Georgi:1974sy}
H.~Georgi and S.~L. Glashow, ``Unity of all elementary particle forces,'' {\em
  Phys. Rev. Lett.} {\bf 32} (1974)
438--441.

\bibitem{Pati:1974yy}
J.~C. Pati and A.~Salam, ``Lepton number as the fourth color,'' {\em Phys.
  Rev.} {\bf D10} (1974)
275--289.

\bibitem{Fritzsch:1974nn}
H.~Fritzsch and P.~Minkowski, ``Unified interactions of leptons and hadrons,''
  {\em Ann. Phys.} {\bf 93} (1975)
193--266.

\bibitem{Slansky:1981yr}
R.~Slansky, ``Group theory for unified model building,'' {\em Phys. Rept.} {\bf
  79} (1981)
1--128.

\bibitem{Asaka:2001eh}
T.~Asaka, W.~Buchm{\"{u}}ller, and L.~Covi, ``Gauge unification in six
  dimensions,'' {\em Phys. Lett.} {\bf B523} (2001) 199--204,
\href{http://www.arXiv.org/abs/hep-ph/0108021}{{\tt hep-ph/0108021}}.

\bibitem{Dimopoulos:1991au}
S.~Dimopoulos, S.~A. Raby, and F.~Wilczek, ``Unification of couplings,'' {\em
  Phys. Today} {\bf 44N10} (1991)
25--33.

\bibitem{Nilles:2004ej}
H.~P. Nilles, ``Five golden rules for superstring phenomenology,''
\href{http://www.arXiv.org/abs/hep-th/0410160}{{\tt hep-th/0410160}}.

\bibitem{Esmaeili::2001ab}
H.~Esmaeili, N.~Mahdavi-Amiri, and E.~E.~Spedicato, ``A class of abs algorithms
  for diophantine linear systems,'' {\em Numer.Math.} {\bf 90} (2001)
101--115.

\bibitem{LLLF82ab}
A.~K. Lenstra, H.~W. Lenstra, and L.~Lov\'asz, ``Factoring polynomials with
  rational coefficients,'' {\em Math. Ann.} {\bf 261} (1982) 515--534.

\bibitem{NTL06}
V.~Shoup, ``{\sc NTL} 5.4,'' {A Library for doing Number Theory}, Courant
  Institute, New York University, 2006.
\newblock {\tt http://shoup.net/ntl}.

\bibitem{Buchberger:1965ab}
B.~Buchberger, {\em Ein Algorithmus zum Auffinden der Basiselemente des
  Restklassenrings nach einem nulldimensionalen Polynomideal}.
\newblock PhD thesis, Universitaet Innsbruck, 1965.

\bibitem{GPS05}
G.-M. Greuel, G.~Pfister, and H.~Sch\"onemann, ``{\sc Singular} 3.0,'' {A
  Computer Algebra System for Polynomial Computations}, Centre for Computer
  Algebra, University of Kaiserslautern, 2005.
\newblock {\tt http://www.singular.uni-kl.de}.

\bibitem{DiFrancesco:1997nk}
P.~Di~Francesco, P.~Mathieu, and D.~Senechal, {\em {Conformal field theory}}.
\newblock Springer, New York, USA, 1997.

\bibitem{NAG}
Numerical Algorithms Group, {\em {NAG C Library}}, 2004.
\newblock \verb+(http://www.nag.co.uk)+.

\bibitem{minilandscape-II}
O.~Lebedev, H.-P. Nilles, S.~Raby, S.~Ramos-Sanchez, M.~Ratz, P.~Vaudrevange,
  and A.~Wingerter, {\em in preparation}.

\bibitem{Dienes:1995sq}
K.~R. Dienes, A.~E. Faraggi, and J.~March-Russell, ``String unification, higher
  level gauge symmetries, and exotic hypercharge normalizations,'' {\em Nucl.
  Phys.} {\bf B467} (1996) 44--99,
\href{http://www.arXiv.org/abs/hep-th/9510223}{{\tt hep-th/9510223}}.

\bibitem{Hall:2001pg}
L.~J. Hall and Y.~Nomura, ``Gauge unification in higher dimensions,'' {\em
  Phys. Rev.} {\bf D64} (2001) 055003,
\href{http://www.arXiv.org/abs/hep-ph/0103125}{{\tt hep-ph/0103125}}.

\bibitem{Contino:2001si}
R.~Contino, L.~Pilo, R.~Rattazzi, and E.~Trincherini, ``Running and matching
  from 5 to 4 dimensions,'' {\em Nucl. Phys.} {\bf B622} (2002) 227--239,
\href{http://www.arXiv.org/abs/hep-ph/0108102}{{\tt hep-ph/0108102}}.

\bibitem{Dienes:1998vh}
K.~R. Dienes, E.~Dudas, and T.~Gherghetta, ``Extra spacetime dimensions and
  unification,'' {\em Phys. Lett.} {\bf B436} (1998) 55--65,
\href{http://www.arXiv.org/abs/hep-ph/9803466}{{\tt hep-ph/9803466}}.

\bibitem{Kim:2002im}
H.~D. Kim and S.~Raby, ``Unification in 5d so(10),'' {\em JHEP} {\bf 01} (2003)
  056,
\href{http://www.arXiv.org/abs/hep-ph/0212348}{{\tt hep-ph/0212348}}.

\bibitem{Hebecker:2004ce}
A.~Hebecker and M.~Trapletti, ``Gauge unification in highly anisotropic string
  compactifications,'' {\em Nucl. Phys.} {\bf B713} (2005) 173--203,
\href{http://www.arXiv.org/abs/hep-th/0411131}{{\tt hep-th/0411131}}.

\bibitem{Ibanez:1992hc}
L.~E. Ibanez and D.~Lust, ``Duality anomaly cancellation, minimal string
  unification and the effective low-energy lagrangian of 4-d strings,'' {\em
  Nucl. Phys.} {\bf B382} (1992) 305--364,
\href{http://www.arXiv.org/abs/hep-th/9202046}{{\tt hep-th/9202046}}.

\bibitem{Giedt:2001zw}
J.~Giedt, ``Spectra in standard-like z(3) orbifold models,'' {\em Annals Phys.}
  {\bf 297} (2002) 67--126,
\href{http://www.arXiv.org/abs/hep-th/0108244}{{\tt hep-th/0108244}}.

\bibitem{Bauer:2000cp}
C.~Bauer, A.~Frink, and R.~Kreckel, ``Introduction to the {\tt {g}i{n}a{c}}
  framework for symbolic computation within the {\tt {c}++} programming
  language,''
\href{http://www.arXiv.org/abs/cs.sc/0004015}{{\tt cs.sc/0004015}}.

\bibitem{Adams:1996groebner}
W.~Adams and P.~Loustaunau, {\em {An Introduction to Gr\"obner Bases.}}
\newblock AMS, Providence, RI, 1996.

\bibitem{Becker:1993groebner}
T.~Becker and V.~Weisspfenning, {\em {Gr\"obner Bases - A Computational
  Approach to Commutative Algebra}}.
\newblock Springer, New York, 1993.

\bibitem{Gray:2006gn}
J.~Gray, Y.-H. He, and A.~Lukas, ``Algorithmic algebraic geometry and flux
  vacua,'' {\em JHEP} {\bf 09} (2006) 031,
\href{http://www.arXiv.org/abs/hep-th/0606122}{{\tt hep-th/0606122}}.

\bibitem{Green:1987sp}
M.~B. Green, J.~H. Schwarz, and E.~Witten, {\em {Superstring Theory. Vol. 1:
  Introduction}}.
\newblock University Press, Cambridge, UK, 1987.

\end{thebibliography}\endgroup
